\documentclass[journal]{IEEEtran}
\ifCLASSINFOpdf
\else
\fi
\usepackage[noend]{algpseudocode}
\usepackage{algorithmicx,algorithm}
\usepackage{cite}
\usepackage{multirow}
\usepackage{amsmath}  
\usepackage{float}
\usepackage{booktabs} 

\usepackage{graphicx}
\usepackage{subfigure}
\usepackage{bm}
\usepackage{lineno}
\usepackage{url}
\usepackage{tikz}
\usepackage{verbatim}
\usetikzlibrary{positioning}
\usepackage{calc}
\usepackage{amsfonts}
\usepackage{rotating}

\begin{document}
\flushbottom
%
\title{Real-Time Steganalysis for Stream Media Based on Multi-channel Convolutional Sliding Windows}
%
%
%

\author{Zhongliang~Yang*,
        Hao~Yang*,
        Yuting~Hu,
        Yongfeng~Huang,
        and~Yu-Jin~Zhang

\thanks{*These authors contributed equally to this work.} 
\thanks{Z. Yang, H.Yang, Y. Hu, Y. Huang and Y. Zhang are with the Department of Electronic Engineering, Tsinghua University, Beijing, 100084, China. E-mail: yangzl15@mails.tsinghua.edu.cn}} 

\maketitle

\begin{abstract}

With the development and popularization of network technology, communication using Voice over IP (VoIP) has become an increasingly common means of information transmission in people's daily lives. With these emerging communication channels, more and more VoIP-based steganography methods have appeared in recent years, which pose a great threat to the security of cyberspace. Previous VoIP steganalysis methods face great challenges in detecting speech signals at low embedding rates, and they are also generally difficult to perform real-time detection, making them hard to truly maintain cyberspace security. To solve these two challenges, in this paper, combined with the sliding window detection algorithm and Convolutional Neural Network (CNN), we propose a real-time VoIP steganalysis method which based on multi-channel convolutional sliding windows (CSW). In order to analyze the correlations between frames and different neighborhood frames in a VoIP signal, we define multi-channel sliding detection windows. Within each sliding window, we design two feature extraction channels which contain multiple convolutional layers with multiple convolution kernels each layer to extract correlation features of the input signal. Then based on these extracted features, we use a forward fully connected network for feature fusion. Finally, by analyzing the statistical distribution of these features, the discriminator will determine whether the input speech signal contains covert information or not. We designed several experiments to test the proposed model's detection ability under various conditions, including different embedding rates, different speech length, etc. Experimental results showed that the proposed model outperforms all the previous methods, especially in the case of low embedding rate, which showed state-of-the-art performance. In addition, we also tested the detection efficiency of the proposed model, and the results showed that it can achieve almost real-time detection of VoIP speech signals. 

\end{abstract}


%
\IEEEpeerreviewmaketitle

\section{Introduction}
%
%
%
%

Concealment system is one of the three basic information security systems in cyberspace \cite{shannon1949communication}, its biggest characteristic is the strong concealment of information. When the other two information security systems, which are encryption system and privacy system, ensure information security, they also expose the existence and importance of information, making it more vulnerable to be attacked by interception and cracking\cite{Bernaille2007Early}. While for the concealment system, it uses various carriers to embed secret information and then transmits them through public channels, in which way the system hides the existence of secret information to achieve the purpose of not being easily suspected and attacked\cite{Simmons1984The}. However, strong concealment can also be used by hackers, terrorists, and other law breakers for malicious intentions. Hence, designing an automatic steganography detection method becomes an increasingly promising and challenging task.

Usually we can model a concealment system as a ``Prisoners' Problem"\cite{Simmons1984The}. In this model, Alice needs to transmit the secret message $m$, which from the secret message space $\mathcal{M}$, to Bob. Alice selects a suitable cover $c$ from the cover space $\mathcal{C}$ and embeds the secret message $m$ into the cover $c$ under the guidance of the hidden key $k_A$, which is from the key space $K$. The cover $c$ becomes a stego carrier $s$ after embedding the covert information $m$, and large number of stego carriers constitute the hidden space $\mathcal{S}$. The information embedding process can be expressed by the embedding function $f$, that is:

\begin{equation}
Emb: \mathcal{C} \times \mathcal{K} \times \mathcal{M} \to \mathcal{S}, f(c,k_A,m) = s.
\end{equation}

\noindent Bob needs to extract the secret message $m$ from the received stego carrier $s$ under the guidance of key $k_B$. The extraction process can be expressed by the extraction function $g$, namely:

\begin{equation}
Ext: \mathcal{S} \times \mathcal{K} \to \mathcal{M}, g(s,k_B) = m.
\end{equation}

\noindent In order to ensure the concealment of secret information, it is usually required that the elements in $\mathcal{S}$ and $\mathcal{C}$ are exactly the same, that is $\mathcal{S} = \mathcal{C}$. But generally speaking, this mapping function will affect the probability distributions, named $P_{\mathcal{C}}$ and $P_{\mathcal{S}}$. For Alice and Bob, their main purpose is to ensure the successful transmission of information without arousing Eve's suspicion, so they need to reduce the difference in statistical distribution of carriers before and after steganography as much as possible, that is:

\begin{equation}
d_f(P_{\mathcal{C}},P_{\mathcal{S}}) \leq \varepsilon .
\end{equation}

\noindent While for Eve, her task is to accurately determine whether the carrier contains hidden information, so she needs to find the difference as much as possible in the statistical distribution of the carrier before and after steganography. 

There are various media forms of carrier that can be used for information hiding, including image\cite{fridrich2009steganography}, audio\cite{yang2017sudoku,huang2011steganography}, text\cite{yang2018rnn,yang2018rits} and so on\cite{johnson2008detection}. In recent years, with the popularity and development of the Internet, communication based on streaming media has been greatly developed. Streaming media refers to continuous time-based media that uses streaming technology in the Internet, such as audio, video, or multimedia files. Voice over IP (VoIP) \cite{Goode2002Voice} is one of the most popular streaming communication service in the Internet. Therefore, with these emerging communication channels, more and more VoIP-based covert communication systems have appeared in recent years \cite{hamdaqa2011relack,tian2009adaptive,xu2011adaptive,ballesteros2012highly,huang2011steganography,tian2014improving,huang2012steganography,yang2017sudoku}. 

Due to the transient and real-time nature of streaming media, steganography methods based on streaming media and static media are very different. For example, methods based on transform domain and spread spectrum are widely used in static audio steganography \cite{kaur2015enhanced,ghasemzadeh2015toward}, however their complexity and time consuming make them not suitable for hiding information in streaming media. A typical streaming media can be represented by Figure 1. Each stream media package contains four data fields: audio and video data (which are usually compressed and coded), IP header, UDP header, RTP header. All of these areas can be used to embed secret information. Therefore, the information hiding technology based on streaming media can be roughly divided into the following two categories: information hiding based on protocol headers \cite{ahsan2002covert,lubacz2014principles} and information hiding based on payloads \cite{wang2007information,huang2012steganography,aoki2006band,chen2001quantization,xiao2008approach,tian2014improving}. For the steganographic methods based on protocol headers, the secret information is mainly embedded into areas of the protocol header which are not commonly used \cite{ahsan2002covert}. This type of methods is simple and easy to implement, but such methods have a low hidden capacity and may cause a great impact on the quality of internet service.

\begin{figure}[ht]
\centering
\includegraphics[width=\linewidth]{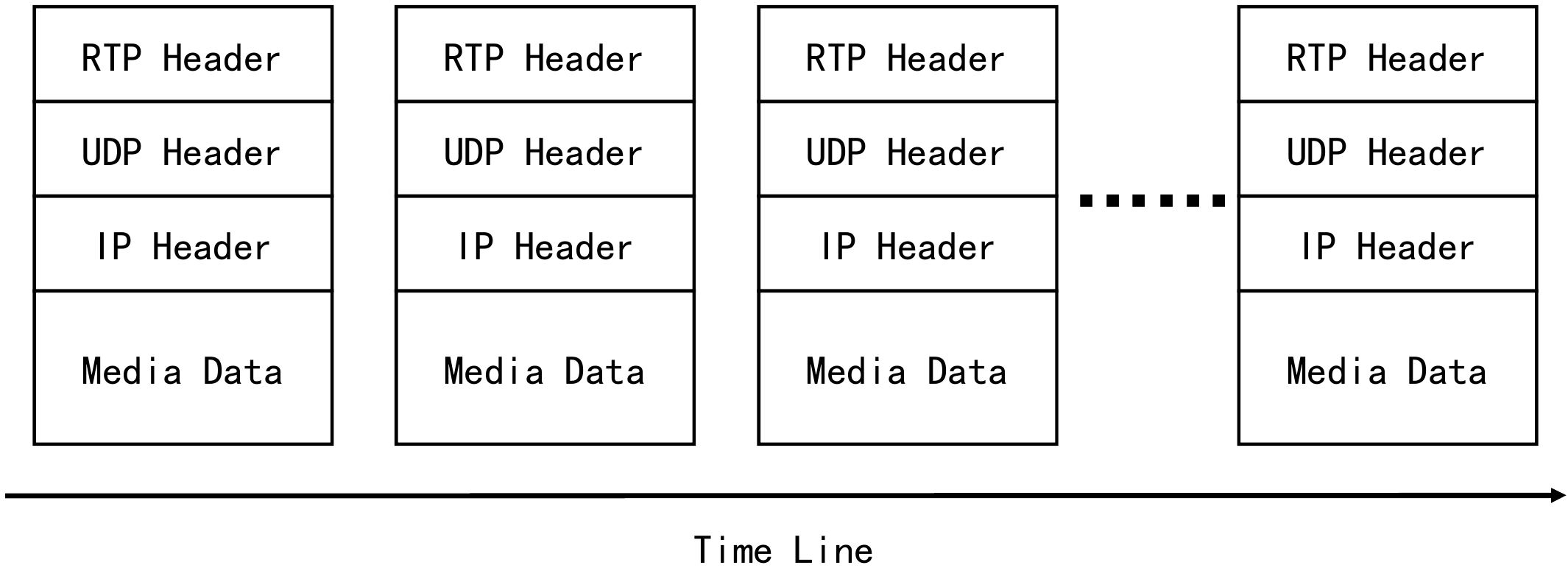}
\caption{A typical data structure for streaming media.}
\label{fig:1}
\end{figure}

At present, information hiding based on payloads are the most common ways for streaming media steganography \cite{wang2007information,huang2012steganography,aoki2006band,chen2001quantization,xiao2008approach,tian2014improving}. They realize information hiding by modifying redundant information of payloads in streaming media, like speech in VoIP. Currently, according to different communication coding standards, the steganographic methods based on payloads in VoIP can be divided into three big families. The first type of methods is mainly for the high-rate Pulse Code Modulation (PCM) speech coding standard G.711. The Least Significant Bits (LSB) algorithm is widely used in this type of steganographic methods \cite{wang2007information,aoki2006band}. They mainly utilize the insensitivity of the human perception system to noise, and replace the LSB of the carrier with the secret information to realize information embedding. The second and third type of methods are mainly for low-bit rate compressed speech coding standards, like G.729 and G.723.1. The second type of methods mainly use the uncertainty of pitch period prediction, and fine-tunes the results of the pitch prediction to achieve the purpose of information hiding \cite{huang2012steganography}. The third type of methods realize information hiding mainly by introducing Quantization Index Modulation (QIM) \cite{chen2001quantization} algorithm to segment and encode the codebook in the process of speech quantization \cite{xiao2008approach,tian2014improving}. With the popularity of the Internet and the widespread use of VoIP, these covert communication methods pose an increasingly serious threat to cyberspace security. Therefore, it is of great value and significance to study the steganalysis method of VoIP and realize fast and high-performance detection of real-time speech stream signals for protecting cyberspace and public security.

Compared with steganalysis methods for static carrier, the steganalysis methods for streaming media are more demanding and thus more challenging. Firstly, since the signal streams are transmitted online in real time, the detection algorithm should also be efficient enough. This involves two requirements. On the one hand, the steganalysis model needs to be able to perform high-performance detection on a speech signal which is as short as possible. Therefore, once Alice and Bob are found to be transmitting secret information, Eve can terminate the communication within the shortest time after they establish the communication. On the other hand, the steganalysis model needs to have a sufficiently fast judgment ability. That is, when inputted a speech signal, it is required to complete the process of feature extraction, analysis and judgment in the shortest time. Secondly, a very important characteristic of covert communication based on VoIP is that, for both communicator parties, the capacity of the carrier is completely controllable and can be expanded arbitrarily. This means that Alice and Bob can spread the secret information that needs to be transmitted in a sufficiently long speech signal to achieve a covert communication with low embedding rate. Therefore, from the perspective of steganalysis, achieve high efficiency and high performance detection of streaming media with low embedding rate has always been an important research goal in the field of VoIP steganalysis.

According to the different information hiding regions, steganalysis methods based on streaming media can also be divided into two categories, which are steganalysis based on network protocols \cite{pelaez2009using} and steganalysis based on payloads \cite{dittmann2005steganography,kraetzer2007pros,huang2011detection,lin2018rnn}. Steganalysis based on network protocols is relatively easier, because each area of the network protocol is carefully designed and clearly defined, the content of each area has obvious statistical characteristics. Once secret information is embedded in some domains of the protocol, the changes can be easily detected \cite{pelaez2009using}. For steganalysis based on payloads, since the corresponding steganographic methods require modification of the carrier to embed information, it is essentially similar to add noise to the carrier, and thus will almost certainly affect the statistical distribution of the carrier in some ways, making it gradually unsatisfactory with formula (3). Therefore, the corresponding steganalysis methods usually analyze the statistical characteristics of the carrier, such as Mel-frequency features \cite{kraetzer2007mel}, codewords correlations \cite{li2012detection,li2017steganalysis,Lin2018Rnn} and so on \cite{dittmann2005steganography,huang2011detection}, by manual construction or model self-learning, and analyze the difference of statistical distribution of these features before and after steganography, then determine whether the inputted VoIP speech contains hidden information. These methods are usually difficult to balance detection efficiency and detection accuracy. Some of them spend a large amount of time on feature extraction and analysis in order to obtain a relatively high detection accuracy, making them difficult to meet the real-time detection requirements \cite{li2017steganalysis}. In addition, some other models may achieve high detection performance at high embedding rate, but when faced with VoIP speech signals with low embedding rate, the detection performance is unsatisfactory \cite{li2012detection}.

In order to solve the two major challenges in the field of VoIP steganalysis, namely: high performance and real-time detection for low embedded rate speech signals, in this paper, combined with the sliding window detection algorithm and Convolutional Neural Network (CNN), we propose a real-time VoIP steganalysis method which based on multi-channel convolutional sliding windows (CSW). It uses multi-channel sliding detection windows to extract correlation features between frames and different neighborhood frames in a VoIP signal. Within each sliding window, we design two feature extraction channels to extract both low-leavel features and high-level features of the input signal. We disigned a large number of experiments to verify our model in many aspects. Experimental results showed that the proposed model outperforms all previous methods, especially for low-embedded VoIP speech signals, and achieved state-of-the-art performance. 

In the remainder of this paper, Section II introduces related work, including QIM-based VoIP Steganography and Speech Steganalysis. Section III introduces the detailed explanation of the proposed method. The following part, Section IV presents the experimental evaluation results and gives a comprehensive discussion. Finally, conclusions are drawn in Section V.

\section{Related Work}

\subsection{QIM-based VoIP Steganography}

To reduce bandwidth usage, VoIP signals are typically first compressed at low-bit rate and then transmitted. Since speech signals are generated by organs in respiratory tract. The organs involved are lung, glottis, and vocal track. When passing through glottis, the exhaled breath from lung would turn to a periodic excitation signal. The excitation signals would then go through vocal track. We can divide vocal track into cascaded segments, whose functions can be modeled as one pole filters. Low-rate speech coding standards used by VoIP such as G.729 and G.723 are based on the Linear Predictive Coding (LPC) technique, which uses the LPC filter to analyze and synthesize acoustic signals in the encoding and the decoding processes. The LPC filter can be expressed as:

\begin{equation}
\label{LPC}
{\text{H(z) = }}\frac{1}{{A(z)}} = \frac{1}{{1 - \sum\nolimits_{i = 1}^n {{a_i}{z^{ - i}}} }},
\end{equation}

\noindent where ${a_i}$ is the (quantized) i-th order coefficient of LPC filter. 

The G.729 and G.723 standards first solve the optimal LPC prediction coefficients for each frame, and then the LPC coefficients are converted into Line Spectrum Frequency (LSF) coefficients. G.729 and G.723 finally use three codewords $(a_1, a_2, a_3)$ to quantize the LSFs. Each codeword $a_i (i = 1, 2, 3)$ has a corresponding codebook $L_i (i = 1, 2, 3)$ whose codeword space is $\{ v_i^1, v_i^2, ..., v_i^{|L_i|} \}$, where $v_i^k$ represents the $k$-th codeword of the codebook $L_i$, and $|L_i|$ is the number of codewords in $L_i$. When quantizing, they select an optimal codeword from the codebook $L_i$. 

The Quantization Index Modulation (QIM) algorithm was firstly proposed by Chen and Wornell \emph{et al.} \cite{chen2001quantization}. They hid secret information by modifying the quantization vector in the media encoding process, thus QIM steganography could be conducted while quantizing LSPs. According to the basic principle of the QIM algorithm, in the vector quantization stage of speech coding, the codebook $C$ is divided into $N$ sub-codebooks, that is

\begin{equation}
\begin{aligned}
&C = \bigcup_{i=1}^{N}C_i,\\
&s.t. \quad \forall i,j\in [1,N], i \neq j,\; C_i\bigcap C_j = \emptyset.
\end{aligned}
\end{equation}

\noindent The QIM-based VoIP steganography encodes each sub-codebook, and then the input vector can be quantized into $N$ different sub-codebook $C_i$ according to the secret information. Usually for QIM-based steganography, the maximum embedding capacity is $\lfloor log_2^{N}\rfloor$ bits. The basic principle of QIM is shown in Figure 2, in which the codebook is divided into four sub-codebooks, so that each quantization can be embedded with $\lfloor log_2^{4}\rfloor = 2$ bits of information.

\begin{figure}[ht]
\centering
\includegraphics[width=\linewidth]{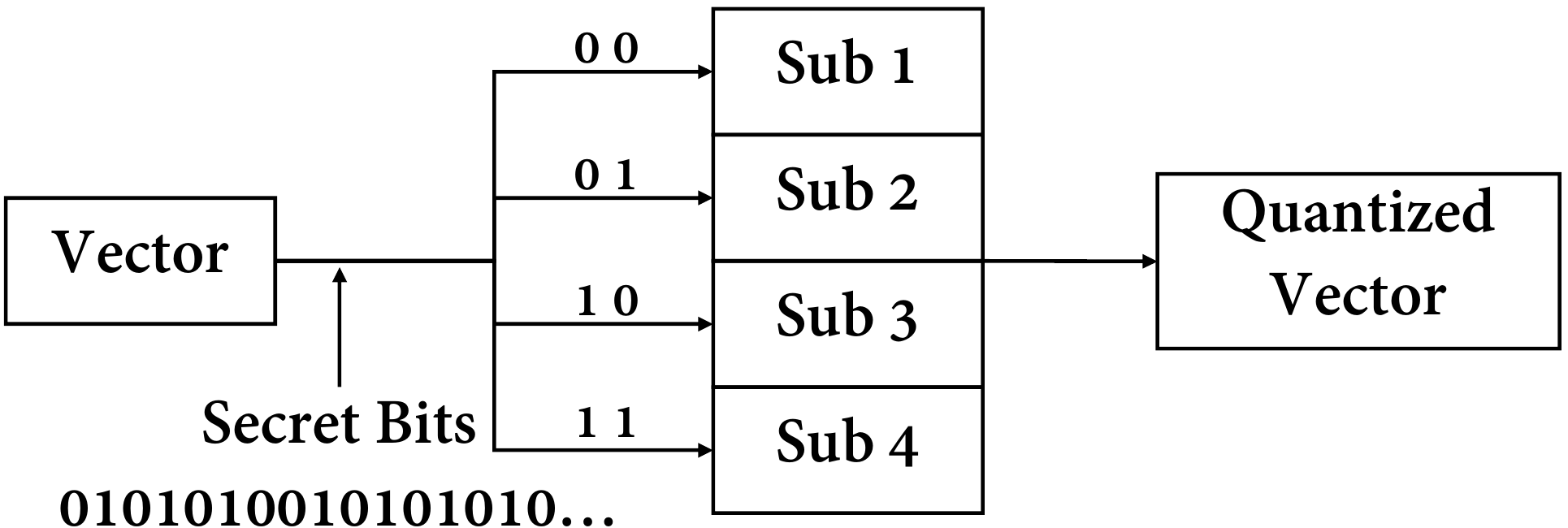}
\caption{The basic principle of Quantization Index Modulation (QIM) algorithm.}
\label{fig:3}
\end{figure}

The key point of the QIM steganography is how to effectively divide the original codebook $L$ into multiple sub-codebooks. To optimize the division of codebooks, Xiao \emph{et al.} \cite{xiao2008approach} proposed a Complementary Neighbor Vertices (CNV) algorithm. This method can optimize the upper limit of quantization distortion after embedding secret information. In the experimental part of this paper, we mainly use the CNV-QIM steganography algorithm as the detection target to test the performance of the proposed method, while the proposed method can also be directly used to detect other QIM-based VoIP steganography methods.

\subsection{Speech Steganalysis}

There has been much effort in steganalysis of digital audio \cite{liu2009temporal,paulin2016audio,kraetzer2007mel,liu2011derivative}. The most common way is to directly extract statistical features from the audio and conduct classification subsequently. For example, C. Kraetzer \emph{et al.} \cite{kraetzer2007mel} analyzed the statistical distribution of Mel-frequency features of audio. Q. Liu \emph{et al.} \cite{liu2009temporal} analyzed the statistics of the high-frequency spectrum and the Mel-cepstrum coefficients of the second order derivative, and further in \cite{liu2011derivative}, they employed the Mel Frequency Cepstrum Coefficient (MFCC) and Markov transition features from the second-order derivative of the audio signal. Then based on these features, these works used Support Vector Machine (SVM) as classifier to decide whether the inputted speech signal contains hidden information or not. 

With the development of neural network technology, more and more speech steganalysis methods based on neural network have appeared in recent years. For example, C. Paulin \emph{et al.} \cite{paulin2016audio} first extracted mel-frequency cepstral coefficients (MFCCs) features of input audio and then used a deep belief networks (DBN) to classify them. S. Rekik \emph{et al.} \cite{rekik2012autoregressive} extracted the Line Spectrum Frequency (LSF) features from original audio and then used a Time Delay Neural Networks (TDNN) to detect stego-speech. These methods are performed by manually extracting the statistical characteristics of the speech signals and then using the neural network models for analysis and detection.

These above audio steganalysis algorithms are all aimed at static audio, which can not be directly applied to VoIP steganalysis due to the unique characteristics of VoIP speech signals. In recent years, there have been many steganalysis methods for VoIP speech signals. For example, the authors of \cite{kraetzer2007mel} and \cite{kraetzer2007pros} thought that steganography of speech streams may lead to low speech quality, so they used Mel-frequency features of speech streams and SVM to determine whether there was hidden information in speech streams. J. Dittmann \emph{et al.} \cite{dittmann2005steganography} successfully implemented the detection of PCM encoded speech stream steganography in the speech stream by extracting the first-order and second-order statistics of the speech stream. Y. Huang \emph{et al.} \cite{huang2011detection} proposed a method for detecting streaming media information hiding based on sliding window. They sampled the speech in the window by selecting an appropriate time window and used Regular Singular (RS) algorithm \cite{fridrich2001detecting} to determine if a LSB replacement had occurred in the speech stream. S. Li \emph{et al.} \cite{li2012detection} used the Markov model to calculate the transition probabilities between frames and then used SVM to classify these features. Further, in \cite{li2017steganalysis}, they considered the transition probability in frames and obtained a better detection effect. Recently, Z. Lin \emph{et al.} \cite{Lin2018RNN} proposed a method for extracting the correlation between codewords and frames in a speech stream using a two-layer Recurrent Neural Network (RNN) with Long Short Time Memory (LSTM) units and then performing steganalysis. These methods are difficult to balance the detection accuracy and detection efficiency. Some of them spend too much time on feature extraction and analysis to improve detection performance, which reduce the detection efficiency. In addition, some other models do not fully analyze the features in order to improve the detection efficiency, which in turn affects the detection performance. The proposed method will optimize both of these aspects. In the experimental part, we will make a detailed comparison and analysis.

\section{The Proposed Method}

The information-theoretic definition of steganographic security starts with the basic assumption that the cover source can be described by a probability distribution, $P_{\mathcal{C}}$, on the space of all possible cover, $\mathcal{C}$. The value $P_{\mathcal{C}}(\mathcal{B})=\int_\mathcal{B}P_{\mathcal{C}}(X)dX$ is the probability of selecting cover $X \in \mathcal{B} \subset \mathcal{C}$ for hiding a message. For a given stegosystem assuming on its input covers $X\in \mathcal{C}, X\sim P_{\mathcal{C}}$ and messages $m \in \mathcal{M}$, the distribution of stego cover is $P_{\mathcal{S}}$. Any steganalysis can be described by a map $\mathit{F}: \mathbb{R}^d \rightarrow \{0,1\}$, where $\mathit{F} = 0$ means that $x$ is detected as cover, while $\mathit{F} = 1$ means that $x$ is detected as stego. 

Figure 3 shows the overall framework of the proposed VoIP steganalysis method. Our model consists of two parts. For an inputted VoIP speech signal, our model first extracts the speech features using multi-channel convolutional sliding windows. Then, after feature fusion, the discriminator determines whether the input speech contains concealed information by analyzing the difference in statistical distribution of these features.

\begin{figure}[ht]
\centering
\includegraphics[width=\linewidth]{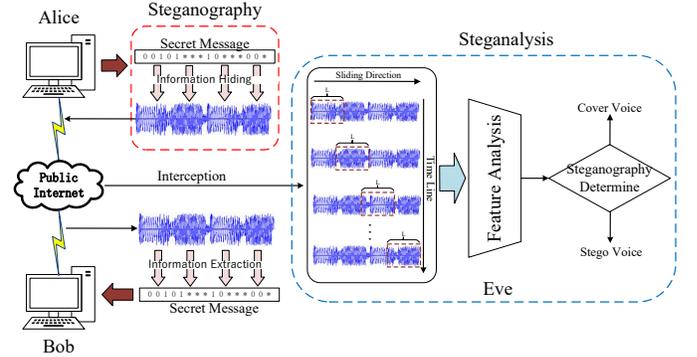}
\caption{The overall framework of the proposed VoIP steganalysis method.}
\label{fig:3}
\end{figure}

\subsection{Feature Analysis}

VoIP speech signals are compressed and encoded sequence signal, so there may have some strong signature correlations between codewords. For an LSF-encoded speech signal which contains $N$ frames, we can express it as:

\begin{equation}
X = [x_1,x_2,...,x_N]=\left[                 
  \begin{array}{cccc}   
    a_{1,1} & a_{1,2} & \cdots\ & a_{1,N}\\  
    a_{2,1} & a_{2,2} & \cdots\ & a_{2,N}\\  
    a_{3,1} & a_{3,2} & \cdots\ & a_{3,N}\\
  \end{array}
\right].
\end{equation}

\noindent Here, $x_i$ denotes the $i$-th frame in the speech segment, and $a_{i,j}$ denotes the $j$-th codeword in the $i$-th frame. When all cordwords are uncorrelated, their appearances are independent. Therefore, we have

\begin{equation}
\begin{aligned}
&P(a_{i,j}, a_{k,l}) = P(a_{i,j}) \cdot P(a_{k,l}),\\
&s.t. \quad \forall i,k\in [1,N];\; j,\;l\in [1,3].
\end{aligned}
\end{equation}

\noindent When the two sides of Equation (7) are not equal, we think there is a correlation between the two cordwords $a_{i,j}$ and $a_{k,l}$. According to Equation (7), we can define two types of codeword correlation: namely intra-frame correlation, which can be expressed as: $P(a_{i,j}, a_{k,l}|i = k\in [1,N],\; j,l\in [1,3])$, and inter-frame correlation, which can be expressed as: $P(a_{i,j}, a_{k,l}|i \neq k,\; j,l\in [1,3])$. Z. Lin \emph{et al.} \cite{Lin2018RNN} further refines the inter-frame correlation into three categories according to the distance between frames: Successive frame Correlation, Cross Frame Correlation, Cross Word Correlation. 

Once we embed additional information into these voice streams, it is possible to influence the correlations between these cordwords and change their statistical distribution. Therefore, some of the previous VoIP steganalysis methods tried to extract these correlation features as a basis for steganography judgment. For example, S. Li \emph{et al.} tried to extract the transition probability feature between inter \cite{li2012detection} and intra \cite{li2017steganalysis} codewords using the Markov model. Z. Lin \emph{et al.} \cite{Lin2018RNN} used a LSTM Neural Nework to extract the temporal correlations of codewords, which achieved currently the best detection performance.

\subsection{Feature Extraction by Convolutional Sliding Windows}

The sliding window algorithm for steganalysis of VoIP speech signal was first proposed by Y. Huang \emph{et al.} \cite{huang2011detection}. They used a fixed-length sliding window to slide over the VoIP data stream. Whitin each each sliding window, they used the Regular Singular (RS) algorithm \cite{fridrich2001detecting} to extract features and then determined if LSB steganography has occurred in the VoIP speech segment. However, they only used a single-channel sliding window, that is, the sliding window had a fixed length, so it could only extract the correlations between each frame and a fixed range of frames. In this proposed model, in order to extract the correlations between each speech frame and the adjacent frames at different distances, we propose a multi-channel sliding window detection method, each channel uses sliding window of different lengths, as shown in Figure 4.

\begin{figure}[ht]
\centering
\includegraphics[width=\linewidth]{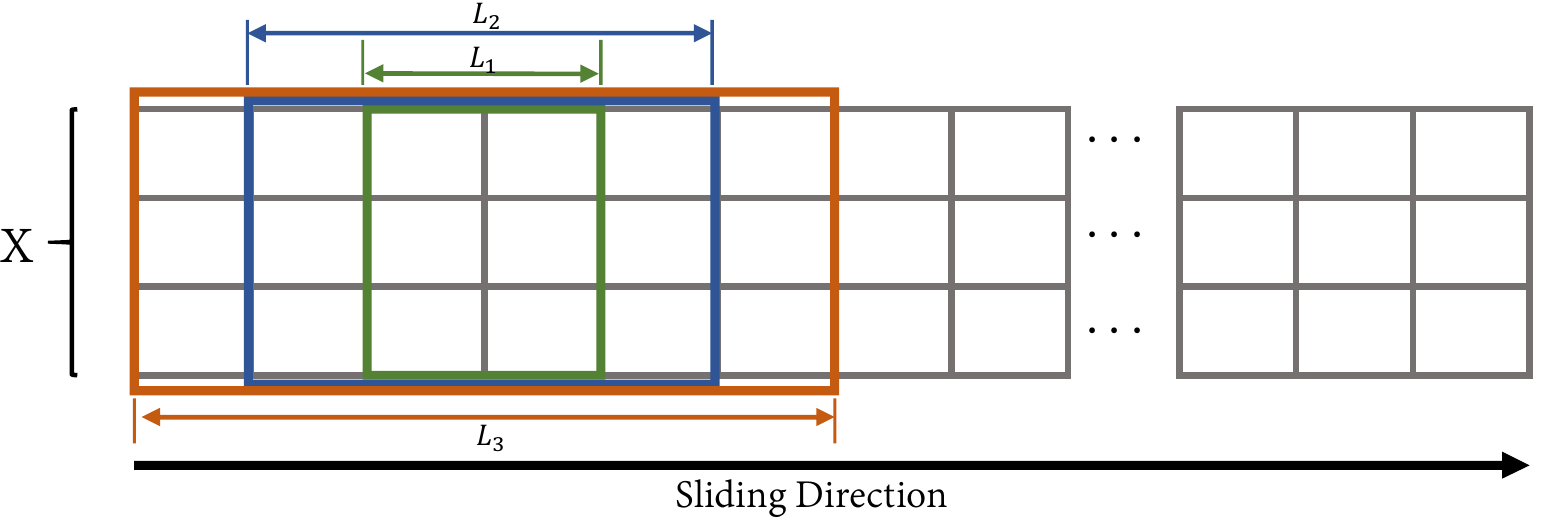}
\caption{We use a multi-channel sliding window detection method, each channel uses sliding window of different lengths.}
\label{fig:3}
\end{figure}

In the past few years, convolutional neural network has made notable progress in fields such as computer vision \cite{krizhevsky2012imagenet,simonyan2014very,yang2017image} and natural language processing \cite{yang2018clinical}. The incremental advancement of CNN is likely to benefit the development of new technology and inventions in other fields. A large number of researches and applications have shown that the convolutional neural network has a powerful ability in feature extractions and expressions \cite{krizhevsky2012imagenet,yang2018clinical}, which does not require hand-designed features but carries out self-learning through plenty of data. Inspired by these works, in the proposed model, for each sliding window, we design a codewords correlation extraction method based on convolutional kernels. Inside each sliding window, we design two feature extraction channels: one is a convolution channel composed of two convolutional layers to extract high-level features, and the other is a skip channel for passing lower-level features. 

Convolution operation is a feature extraction process for the elements in the local region of the input matrix. More specifically, suppose that the width of the $k$-th sliding window is $L_k$, then the convolution kernel of the first convolutional layer can be expressed as $W^k \in \mathbb{R}^{3 \times L_k}$, that is

\begin{equation}
W^k = 
\left[                 
  \begin{array}{cccc}   
    w^k_{1,1} & w^k_{1,2} & \cdots\ & w^k_{1,L_k}\\  
    w^k_{2,1} & w^k_{2,2} & \cdots\ & w^k_{2,L_k}\\  
    w^k_{3,1} & w^k_{3,2} & \cdots\ & w^k_{3,L_k}\\
  \end{array}
\right].
\end{equation}

\noindent When $w^k_{1,1}$ and $a_{1,1}$ coincide, the feature $c^k_{1,1}$ extracted from $x_{1:h}$ by the convolutional kernel can be:

\begin{equation}
c^k_{1,1} = f(\sum_{i=1}^{3}\sum_{j=1}^{l_k} w^k_{i,j}\cdot a_{i,j} + b^k_{i,j}),
\end{equation}

\noindent where the weight $w^k_{i,j}$ denotes the contribution of the $j$-th value in the $i$-th frame, $b^k_{i,j}$ is the bias term and $f$ is a nonlinear function. Here we follow previous works \cite{krizhevsky2012imagenet} and use ReLu function as our nonlinear function, which is defined as

\begin{equation}
y=ReLu(x) = max(0,x).
\end{equation}

It is worth noting that the feature $c^k_{1,1}$ here is only extracted from a single convolution kernel. Previous works have shown that different convolution kernels may extract features from different aspects of the input signal \cite{krizhevsky2012imagenet}. Therefore, in order to extract more comprehensive features of the speech signal in the sliding window, we simultaneously use multiple convolution kernels for feature extraction. Suppose the number of convolution kernels in the first convolutional layer is $u_1$, when $w^k_{1,1}$ and $a_{1,1}$ coincide, the extracted features can be expressed as

\begin{equation}
C^k_1 = [c^k_{1,1},c^k_{1,2},...,c^k_{1,u_1}].
\end{equation}

\begin{figure}[!tp]
\centering
\includegraphics[width=\linewidth]{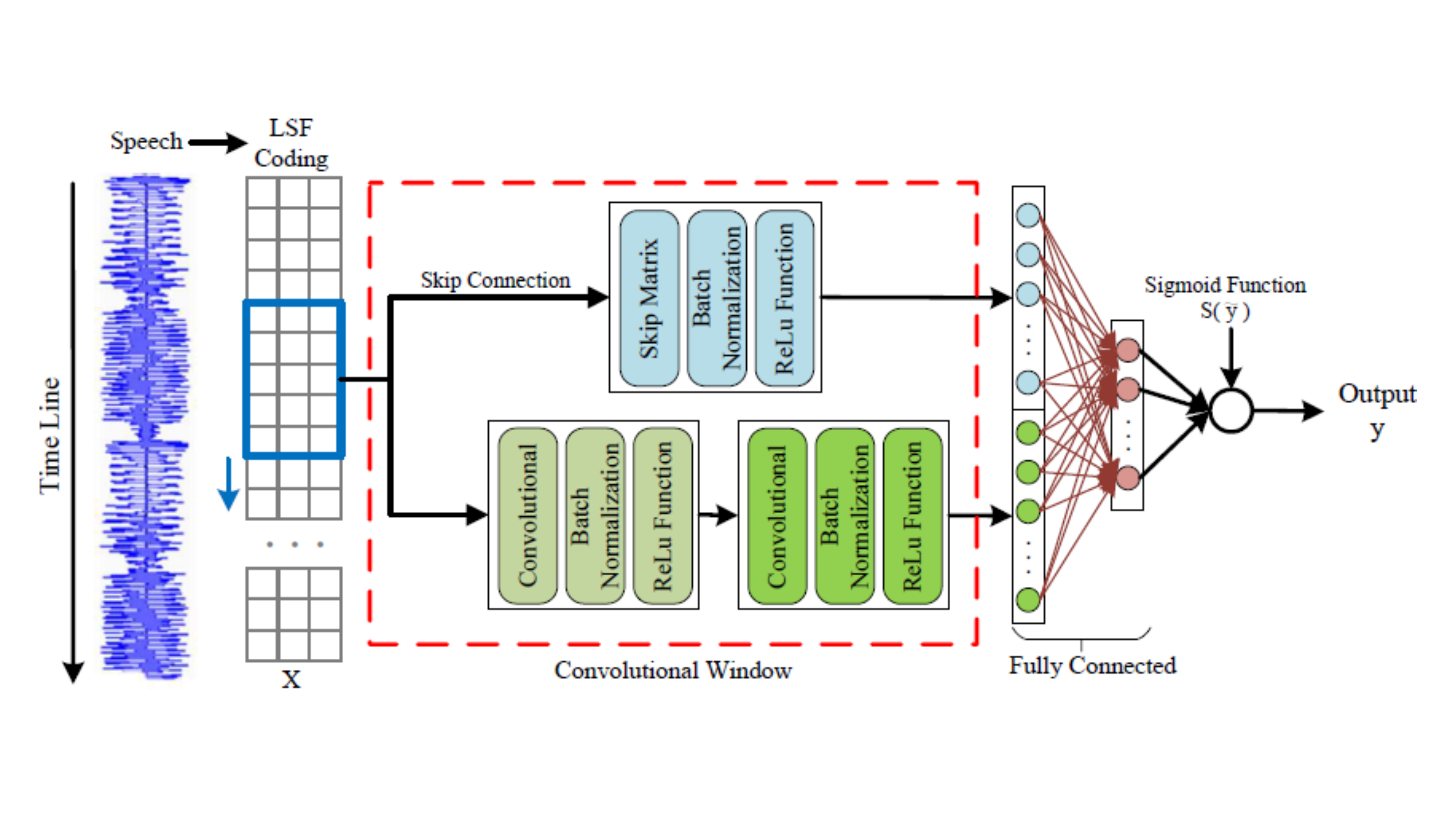}
\caption{The details of the proposed method. For each sliding window, we design two feature extraction channels: one is a convolution channel composed of two convolutional layers to extract high-level features, and the other is a skip channel for passing lower-level features. Finally, we use a forward neural network to fuse the extracted features and finally determine whether the input speech contains hidden information.}
\label{fig:3}
\end{figure}

In order to further extract the codeword correlation features in different regions of the input speech signal, we slide the $k$-th window from the beginning of input speech sample $X$ to the end of it, and calculates the features of each local region. Therefore, after the first convolutional layer, the features we extracted can be expressed as $C^k \in \mathbb{R}^{l_c \times u_1}$, that is

\begin{equation}
C^k = \left[                 
  \begin{array}{cccc}   
    c^k_{1,1} & c^k_{1,2} & \cdots\ & c^k_{1,u_1}\\  
    c^k_{2,1} & c^k_{2,2} & \cdots\ & c^k_{2,u_1}\\  
    \vdots & \vdots & \ddots & \vdots\\
    c^k_{l_c,1} & c^k_{l_c,2} & \cdots\ & c^k_{l_c,u_1}\\
  \end{array}
\right],
\end{equation}

\noindent where $l_c$ represents the number of features extracted by each convolution kernel for the input speech signal. When the sliding step is $T_c$ (which we usually set it to be $1$), the relationship between $l_c$ and the input signal length $N$ is:

\begin{equation}
l_c = \frac{N-L_k+1}{T_c}.
\end{equation}

Previous work on convolutional neural networks has shown that increasing the number of convolutional layers within a certain range is beneficial for extracting higher-level signal features and also for later feature analysis \cite{simonyan2014very}. However, too many layers can lead to over-fitting and reduce detection efficiency. So we add another convolutional layer after the first convolutional layer. The operation of the second convolutional layer and the first convolutional layer are basically the same. It also contains multiple convolution kernels and each has the same width as $C^k$. Then these convolution kernels slid from top to bottom and using convolution calculations to extract high-level features. Therefore, the features extracted by the second convolutional layer can be expressed as $E^k \in \mathbb{R}^{l_e \times u_2}$, that is

\begin{equation}
E^k = \left[                 
  \begin{array}{cccc}   
    e^k_{1,1} & e^k_{1,2} & \cdots\ & e^k_{1,u_2}\\  
    e^k_{2,1} & e^k_{2,2} & \cdots\ & e^k_{2,u_2}\\  
    \vdots & \vdots & \ddots & \vdots\\
    e^k_{l_e,1} & e^k_{l_e,2} & \cdots\ & e^k_{l_e,u_2}\\
  \end{array}
\right].
\end{equation}

\noindent Where $l_e$ represents the number of features extracted by each convolution kernel of the second layer and $u_2$ represents the number of convolution kernels in the second convolutional layer.

In addition, we also think that although the convolutional kernels extract high-level features, they ignore some details of the original signal, which we think are useful for the final analysis. Therefore, referring to the latest works in the field of neural networks \cite{larsson2016fractalnet,huang2017densely}, we add an additional skip connection to pass the low-level features of the original signal. In order to facilitate the feature fusion of the low-level features and the high-level features extracted by the convolutional layer, we define a skip matrix on the skip connection. The skip matrix $S^k$ is mainly for dimensional transformation and simple feature extraction of the original signal, which can be expressed as $S^k \in \mathbb{R}^{l_s \times 3}$:

\begin{equation}
S^k = 
\left[                 
  \begin{array}{cccc}   
    s^k_{1,1} & s^k_{1,2} & \cdots\ & s^k_{1,L_s}\\  
    s^k_{2,1} & s^k_{2,2} & \cdots\ & s^k_{2,L_s}\\  
    s^k_{3,1} & s^k_{3,2} & \cdots\ & s^k_{3,L_s}\\
  \end{array}
\right]^\top.
\end{equation}

\noindent Similarly, the low-level features $F^k$ of the original signal it extracts are

\begin{equation}
F^k = f(S^kX + b_s^k)
\end{equation}

\noindent where the weight $b_s^k$ is the bias term and $f$ is a nonlinear function which is also setted to be ReLu function. 

Finally, in order to enhance the robustness of the model and prevent overfitting, we add batch normalization \cite{ioffe2015batch} after each layer. This component is widely used in the field of neural networks and has been proven to effectively improve the training and testing performance of the network.

\subsection{Feature Fusion and Steganography Determine}

The model shown in Figure 5 can extract the correlation between each frame in the speech signal and the surrounding frames in different ranges. On this basis, we need to fuse the extracted features further and make steganalysis based on the statistical characteristics of these features. Firstly, we use the pooling layer to fuse the features extracted from multiple convolution kernels in a single sliding channels. The pooling layer is widely used in neural network-related models. It can reduce the number of neural network parameters while maintaining the overall distribution of features, which can effectively prevent the model from over-fitting and improve the robustness of the model \cite{krizhevsky2012imagenet,Boureau2010A}. In the proposed model, we conduct a max pooling operation on the feature $E^k$ and $F^k$. For example, for the $i$-th row feature extracted in $E$ and $F$, the outputs are:

\begin{equation}
\begin{aligned}
&P^k = [p^k_1,p^k_2,...,p^k_{l_e}]^\top,\\
&Q^k = [q^k_1,q^k_2,...,q^k_{l_e}]^\top,
\end{aligned}
\end{equation}

\noindent where

\begin{equation}
\begin{aligned}
&p^k_i = max(e^k_{i,1}, e^k_{i,2}, \cdots\, e^k_{i,u_2}),\\
&q^k_i = max(s^k_{i,1}, s^k_{i,2}, \cdots\, s^k_{i,u_2}).
\end{aligned}
\end{equation}

Usually in order to preserve a richer feature distribution, we can use k-max pooling. k-max pooling is a generalisation of the max pooling operator, it returns the subsequence of $k$ maximum values in the input features, instead of the single maximum value. An explanation of max pooling and 2-max pooling have been shown in figure 6.

\begin{figure}[ht]
\centering
\subfigure[max-pooling]{
\begin{minipage}[t]{0.45\linewidth}
\centering
\includegraphics[width=\linewidth]{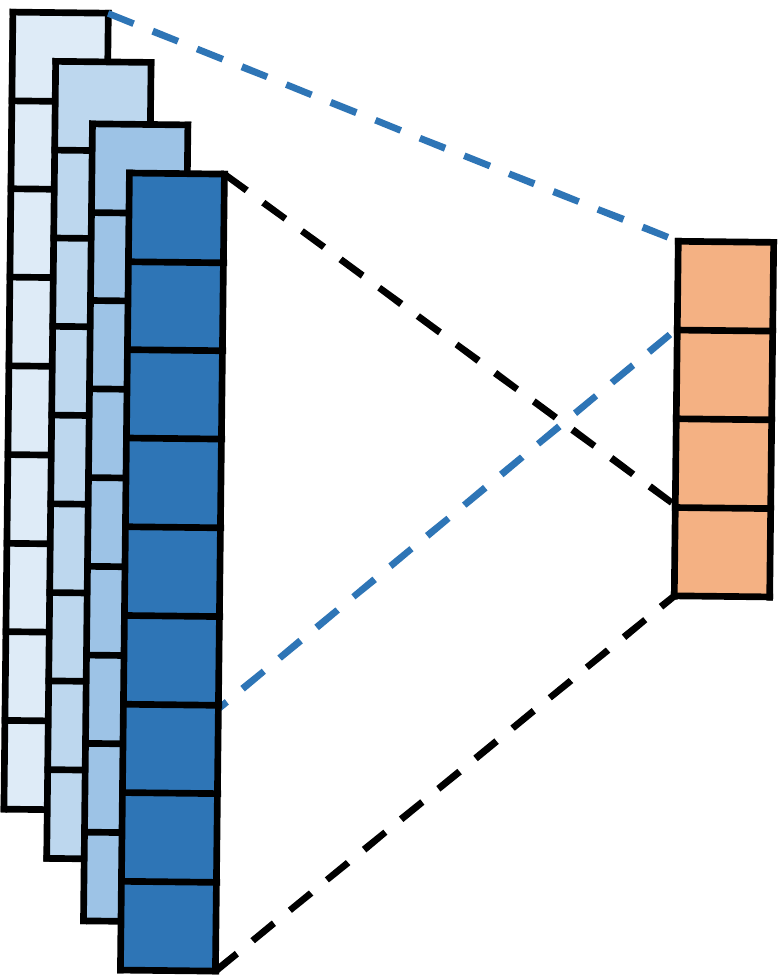}
\end{minipage}%
}%
\subfigure[2-max pooling]{
\begin{minipage}[t]{0.45\linewidth}
\centering
\includegraphics[width=\linewidth]{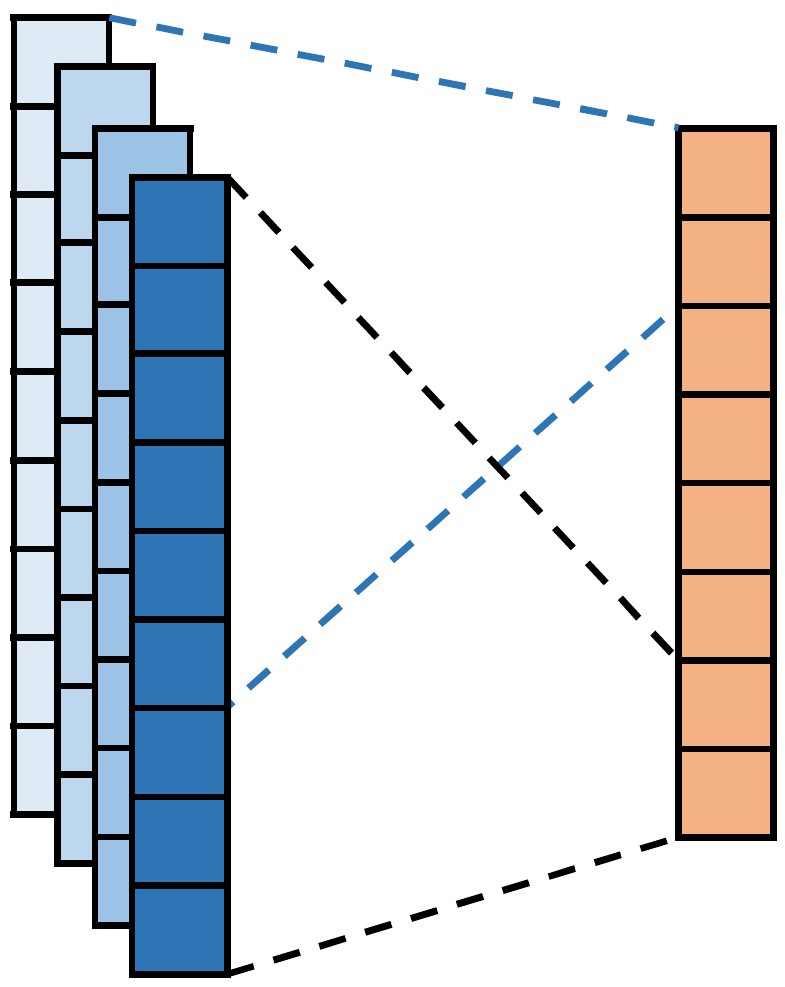}
\end{minipage}%
}%
\centering
\caption{The explanation of max pooling and 2-max pooling.}
\end{figure}

Secondly, we need to fuse the features extracted from different local parts of the input signal by different sliding windows. In order to achieve the fusion of these features, and also to further extract the correlation between the frames in far distance, we add a fully connected forward neural network. To be specifically, we splice the low-level and high-level features of different local regions extracted by a single sliding window, that is, $P^k$ and $Q^k$ are spliced into a complete feature vector $Z^k$, that is:

\begin{equation}
Z^k = [(P^k)^\top;(Q^k)^\top]^\top.
\end{equation}

\noindent Then we connect the features extracted by different sliding channels and get the complete feature expression $Z$ of the input speech signal:

\begin{equation}
Z = [z_1,z_2,...,z_m] = [(Z^1)^\top;(Z^2)^\top;...;(Z^M)^\top]^\top,
\end{equation}

\noindent where $M$ is the number of channels and $m$ is the dimension of the merged feature $Z$.

Then we define a Feature Fusion Matrix $W^{o} \in \mathbb{R}^{{m} \times {h}}$ to fuse the features, where $h$ is the dimension of the fused feature vector $O = [o_1,o_2,...,o_h]$, where:

\begin{equation}
o_{j} = \sum^m_{i=1}w^o_{i,j}\cdot z_i + b^o_{j}.
\end{equation}

We can then use the features collected in $O$ to classify whether the original speech signal contains secret information. A basic idea is to calculate the linear combination of all features. Following previous works \cite{Lin2018RNN,yang2018clinical}, we define the Detection Weight Vector (DWV) as $V$ with a length of $h$, and the linear combination is calculated as

\begin{equation}
\widetilde{y} = \sum^{h}_{i=1}o_{i}\cdot v_{i}+b^o_i.
\end{equation}

\noindent To get normalized output between $[0, 1]$, we send the value through a sigmoid function $S$:

\begin{equation}
S(x) = \frac{1}{1+e^{-x}},
\end{equation}

\noindent and the final output is 

\begin{equation}
y = S(\widetilde{y}) = S(\sum^{h}_{i=1}o_{i}\cdot v_{i}+b^o_i).
\end{equation}

\noindent Finally, we can set a detection threshold (like $0.5$) and then the final detection result can be expressed as

\begin{equation}
X \in 
\left\{
\begin{aligned}
& stego \ speech & (y \geq threshold) \\
& cover\ speech & (y < threshold)
\end{aligned}
\right.
\end{equation}

To determine the parameters in the proposed model, we follow a supervised learning framework. In the process of training, we update network parameters by applying backpropagation algorithm, and the loss function of the whole network consists of two parts, one is the error term and the other is the regularization term, which can be described as:

\begin{equation}
LOSS = - \frac{1}{N}\sum_{N}t_i \cdot log(y_i) + \|W^{o}\|_2,
\end{equation}

\noindent where $N$ is the batch size of VoIP signals. $y_i$ represents the probability that the $i$-th sample is judged to contain covert information, $t_i$ is the actual label of the $i$-th sample. The error term in the loss function calculates the average cross entropy between the predicted probability value and the real label. We hope that through self-learning of the model, prediction error can get smaller and smaller, that is, the prediction results are getting closer to the real label. In order to strengthen the regularization and prevent overfitting, we adopt the dropout mechanism and a constraint on l2-norms of the weight vectors during the training process. Dropout mechanism means that in training process of deep learning network, the neural network unit is temporarily discarded from the network, i.e. set to zero, according to a certain probability. This mechanism has been proved to effectively prevent neural network from overfitting, and significantly improve the model's performance.

\section{Experiments and Analysis}

In this section, we conducted several experiments and compared the results with some state of the art VoIP steganalysis methods to verify the validity of the proposed model. This section starts with an introduction of the dataset used in this work, followed by the structure and the hyper-parameters of the model. Finally, we compare the performance of the proposed model and other state-of-the-art VoIP steganalysis models under different conditions, including different embedding rates, different durations and so on.

\subsection{Dataset Collection and Evaluation Method}

We use the speech dataset\footnote{\url{https://github.com/fjxmlzn/RNN-SM}} published by Z. Lin \emph{et al.} \cite{Lin2018RNN} to train and test the performance of the proposed model. This dataset contains 41 hours of Chinese speech and 72 hours of English speech in PCM format with 16 bits per sample from the Internet. The speech samples are from different male and female speakers. Those speech samples make up the cover speech dataset.

For each sample in this speech dataset, they first use the G.729A standard to encode it and get LSF matrix. Random 01 bit stream is embedded using CNV-QIM steganography proposed in \cite{xiao2008approach}, and these samples make up the stego speeches. Embedding rate is defined as the ratio of the number of embedded bits to the whole embedding capacity. Lower embedding rate indicates fewer changes to the original signal streams. When conducting a\% embedding rate steganography, we embed each frame with a\% probability. In order to test the performance of the proposed model at different embedding rates (especially low embedding rate), we performed 10\%, 20\%, 30\%, ..., 100\% different embedding rates for the speech samples in the dataset. In addition to embedding rate, sample length is another factor that influences detection accuracy. Samples in the speech dataset are cut into 0.1s, 0.2s,....,10s clips. Segments of the same length are successive and nonoverlapped.

For model training, we random choose 80\% of samples, which in the same language and have the same length, from cover speech and stego speech as positive samples and negative samples respectively, and the remaining 20\% for test and validate. For example, for 0.1s clips with 1:1 ratio of cover clips and stego clips, the training set has 2,486,708 samples and the test set contains 155,405 clips.

In the experimental part, we mainly use $Accuracy$ $(Acc)$, $FP$ and $FN$ to measure the detection performance of each model. Here, $Accuracy$ calculates the proportion of true results (both true positives and true negatives) among the total number of cases examined:

\begin{equation}
Accuracy = \frac{TP + TN}{TP + FN + FP + TN}.
\end{equation}

\noindent Where TP (True Positive) represents the number of positive samples that are predicted to be positive by the model, FP (False Positive) indicates the number of negative samples predicted to be positive, FN (False Negative) illustrates the number of positive samples predicted to be negative and TN (True Negative) represents the number of negative samples predicted to be negative.

\subsection{Experimental Setting and Training Details}

Almost all the parameters in the proposed model can be obtained through training, but there are still some hyper-parameters that need to be determined, such as the number of channels, the number of concolutional layers, and so on. Generally, increasing these hyper-parameters will enhance network' representation ability. However, it may also increase the possibility of overfitting and reduce detection performance.

To determine these hyper-parameters, we designed multiple sets of comparative experiments. Table \ref{modelv} shows nine different model settings, whose components are slightly different from the final proposed model (model $a$). Figure 7 shows the detection accuracy of these nine fine-tuning models for 10s' samples with 10\% embedding rate.

\begin{table}[ht]
\renewcommand\arraystretch{1.3}
  \centering
  \caption{A description of the various model settings.}
    \begin{tabular}{c|p{24.5em}}
    \toprule[1.5pt]
    \multicolumn{1}{l|}{\textit{\textbf{ models}}} & \multicolumn{1}{c}{\textit{\textbf{Model Settings}}} \\
    \hline
    \textbf{a} & The full proposed model. \\
    \hline
    \textbf{b} & Remove the skip connection. \\
    \hline
    \textbf{c} & Replace all 2-max-pooling with max pooling.  \\
    \hline
    \textbf{d} & All pooling layer use 2-max pooling.  \\
    \hline
    \textbf{e} & All pooling layer use 3-max pooling.  \\
    \hline
    \textbf{f} & Remove the First convolutional layer. \\
    \hline
    \textbf{g} & Remove the second convolutional layer. \\
    \hline
    \textbf{h} & Set the number of the convolutional layer to 3. \\
    \hline
    \textbf{i} & Set the number of the convolutional layer to 4. \\
    \hline
    \textbf{j} & Change the number of channels of the sliding window to 2. \\
    \bottomrule[1.5pt] 
    \end{tabular}%
  \label{modelv}%
\end{table}%

\begin{figure}[ht]
\centering
\includegraphics[width=\linewidth]{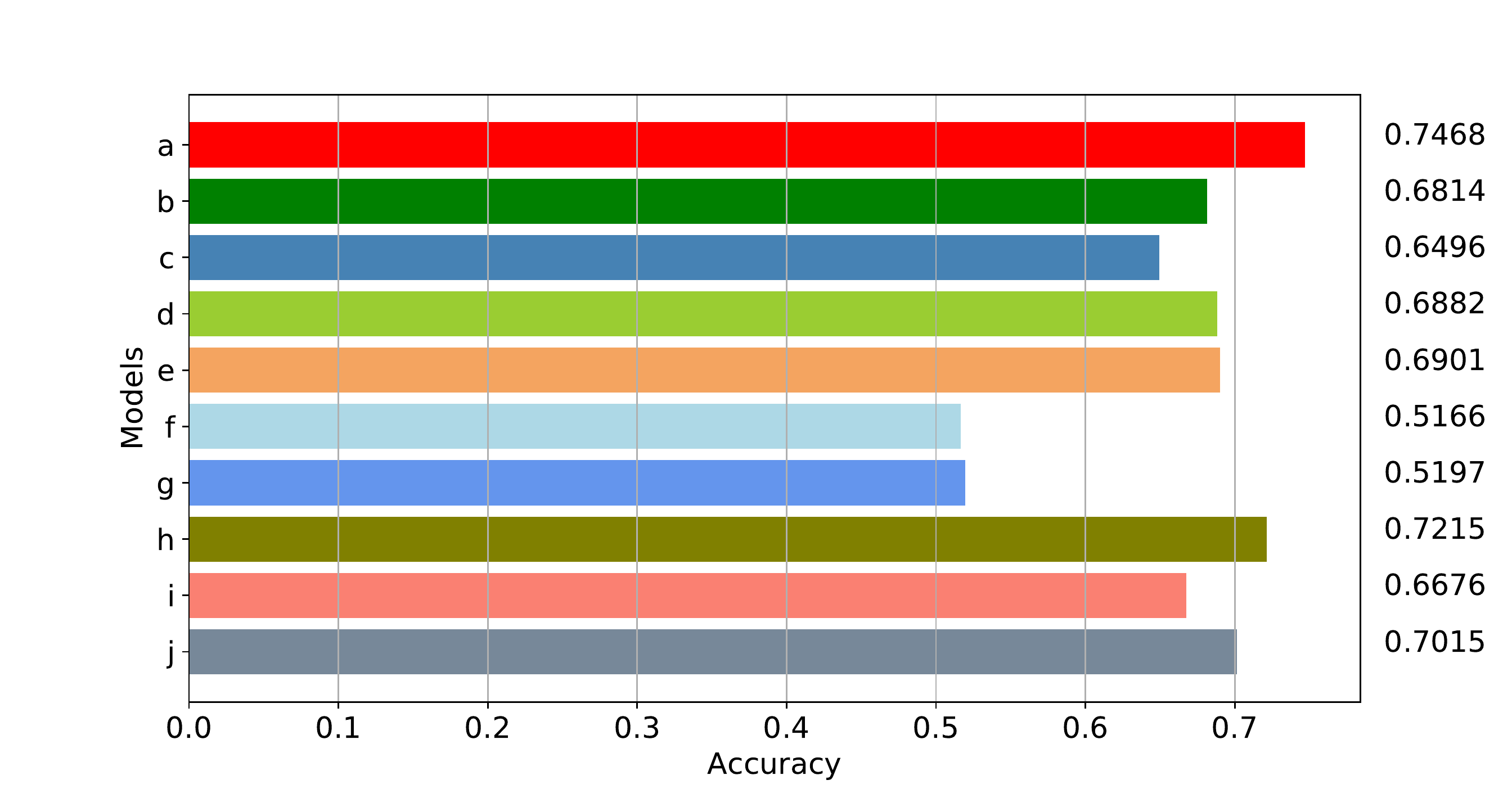}
\caption{The detection accuracy of various models which aim at speech signal with 10\% embedding rate and 10s length.}
\label{fig:3}
\end{figure}

Firstly, we can see from Figure 7 that the convolutional layer has a very significant impact on the detection performance of the entire model. For example, when we compare model $a$ with model $f$ and $g$, we can see that when we delete any convolution layer, the detection performance of the final model is greatly reduced. But when we increase the number of layers of the convolution layer, such as the models $h$ and $i$, it also reduces the detection performance of the model. This is consistent with our previous analysis, that is, it's not the more the number of convolution layers, the better the performance, because more layers will increase the risk of over-fitting and then lower the detection performance. Secondly, when we compare model $a$ and model $b$, we find it necessary to add the skip connection in each sliding window. As we analyzed in model section, the added skip connection in the sliding window is mainly to pass some original features of the input signal. Because we think these original features contain more details and can be helpful for the final steganalysis, which has been proved by the experimental results. Third, comparing the results of model $a$ and model $j$, we find that increasing the number of channels is valuable. Because for the original single channel sliding window detection algorithm, it can only extract the correlation within the fixed neighborhood of each signal frame. However, multi-channel sliding windows can extract the correlation of different neighborhoods around each frame. Therefore, the detection performance of multi-channel sliding windows is better. Finally, we also compare the impact of the pooling operation on the final performance during the feature fusion part. Specifically, we can compare the detection results of model $a$, model $d$ and model $e$. Through multiple comparison experiments, the pooling layer is finally setted as follows: 2-max pooling at the end of convolutional layer and 1-max pooling at the end of skip connected layer.

\begin{table*}[!ht]
\centering
\caption{\label{tab:4}For speech with a length of 10s, the detection performance of each model at different embedding rates.}
\resizebox{\textwidth}{50mm}{
\begin{tabular}{c|c|c|cccccccccc}
\toprule[2pt]
\multirow{2}{*}{Language} &\multirow{2}{*}{Method} &\multirow{2}{*}{Metric} &\multicolumn{10}{|c}{Embedding Rate}\\

 & & &10\% &20\% &30\% &40\% &50\% &60\% &70\% &80\% &90\% &100\%\\
\hline
\multirow{12}{*}{English} &\multirow{3}{*}{IDC \cite{li2012detection}} &Acc (\%) &51.60 & 58.55 & 63.65 & 71.50 & 76.25 & 83.50 & 87.25 & 91.60 & 95.55 & 97.20\\
& &FP (\%) &51.50 &44.70 &39.60 &31.10 &28.30 &19.60 &14.10 &9.00 &4.50 &2.40\\
& &FN (\%) &45.30 &38.20 &33.10 &25.90 &19.20 &13.40 &11.40 &7.80 &4.40 &3.20\\
\cline{2-13}
&\multirow{3}{*}{SS-QCCN \cite{li2017steganalysis}} &Acc (\%) &54.40 &75.45 &92.45 &97.35 &99.15 &99.60 &\textbf{100.00} &\textbf{100.00} &99.95 &\textbf{99.30}\\
& &FP (\%) &55.50 &32.30 &10.80 &4.10 &1.20 &0.60 &0.00 &0.00 &0.00 &0.00\\
& &FN (\%) &35.70 &16.80 &4.30 &1.20 &0.50 &0.20 &0.00 &0.00 &0.10 &1.40\\
\cline{2-13}
&\multirow{3}{*}{RNN-SM \cite{Lin2018RNN}} &Acc (\%) &59.64 &92.44 &94.56 &96.90 &97.76 &98.77 &99.24 &99.71 &99.79 &98.78\\
& &FP (\%) &38.70 &9.18 &5.73 &3.95 &1.98 &2.12 &0.84 &0.31 &0.29 &0.04\\
& &FN (\%) &42.64 &5.94 &5.16 &2.24 &2.50 &0.34 &0.68 &0.27 &0.13 &2.39\\
\cline{2-13}
&\multirow{3}{*}{Ours} &Acc (\%) &\textbf{83.48} &\textbf{94.15} &\textbf{97.76} &\textbf{99.17} &\textbf{99.71} &\textbf{99.91} &99.95 &99.98 &\textbf{100.00} &99.05\\
& &FP (\%) &10.50  &8.49  &2.94  &0.30  &0.25  &0.01  &0.07  &0.03 &0.00  &1.59\\
& &FN (\%) &22.54  &3.21  &1.54  &1.36  &0.33  &0.17  &0.03  &0.01  &0.00  &0.31\\
\hline
\hline
\multirow{12}{*}{Chinese} &\multirow{3}{*}{IDC \cite{li2012detection}} &Acc (\%) &52.75 &59.25 &65.55 &71.40 &78.50 &82.60 &89.15 &93.60 &96.05 &98.05\\
& &FP (\%) &47.30 &45.20 &39.80 &34.10 &26.40 &21.20 &13.00 &8.00 &5.50 &2.10\\
& &FN (\%) &47.20 &36.30 &29.10 &23.10 &16.60 &13.60 &8.70 &4.80 &2.40 &1.80\\
\cline{2-13}
&\multirow{3}{*}{SS-QCCN \cite{li2017steganalysis}} &Acc (\%) &57.35 &75.00 &92.00 &98.25 &99.50 &99.85 &\textbf{100.00} &99.95 &99.90 &\textbf{99.75}\\
& &FP (\%) &45.50 &29.60 &12.10 &3.00 &0.90 &0.20 &0.00 &0.10 &0.00 &0.00\\
& &FN (\%) &39.80 &20.40 &3.90 &0.50 &0.10 &0.10 &0.00 &0.00 &0.20 &0.50\\
\cline{2-13}
&\multirow{3}{*}{RNN-SM \cite{Lin2018RNN}} &Acc (\%) &55.14 &74.19 &90.12 &95.24 &98.05 &98.25 &99.09 &99.51 &99.76 &99.55\\
& &FP (\%) &71.18 &33.27 &10.71 &6.97 &2.07 &1.59 &0.22 &0.52 &0.06 &0.26\\
& &FN (\%) &19.87 &18.66 &9.05 &2.60 &1.83 &1.91 &1.61 &0.45 &0.43 &0.66\\
\cline{2-13}
&\multirow{3}{*}{Ours} &Acc (\%) &\textbf{74.68} &\textbf{92.05} &\textbf{96.58} &\textbf{98.70} &\textbf{99.64} &\textbf{99.87} &99.94 &\textbf{99.98} &\textbf{100.00} &99.51\\
& &FP (\%) &15.21  &13.28 &1.12  &1.57  &0.05  &0.06  &0.03  &0.03  &0.00  &0.07\\
& &FN (\%) &35.43  &2.62  &5.72  &1.03  &0.67  &0.20  &0.09  &0.01  &0.00  &0.91\\
\bottomrule[2pt] 
\end{tabular}}
\end{table*}

\begin{figure*}[!ht]
\centering
\subfigure[English]{
\begin{minipage}[t]{0.50\linewidth}
\centering
\includegraphics[height=7cm,width=\linewidth]{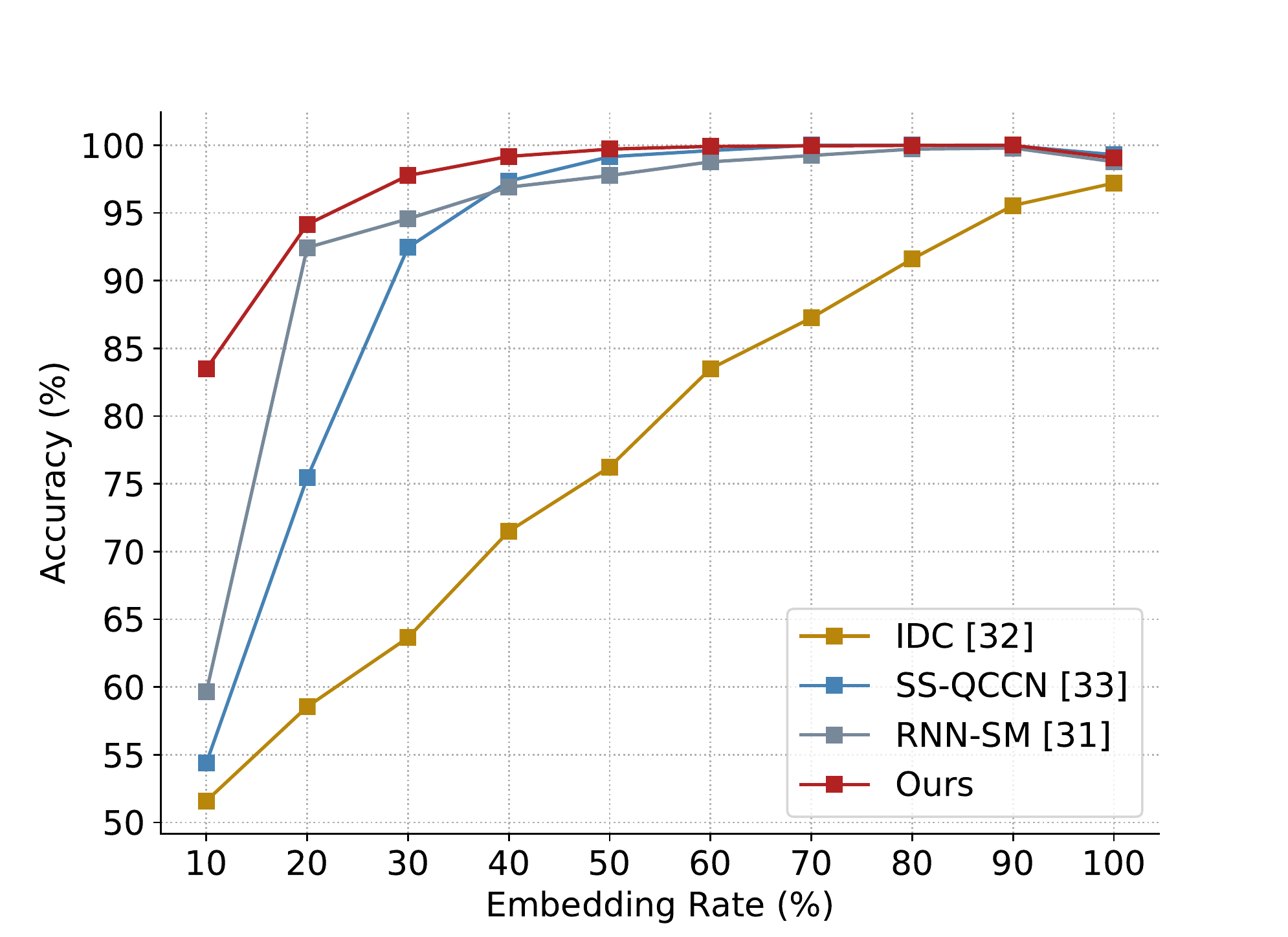}
\end{minipage}%
}%
\subfigure[Chinese]{
\begin{minipage}[t]{0.50\linewidth}
\centering
\includegraphics[height=7cm,width=\linewidth]{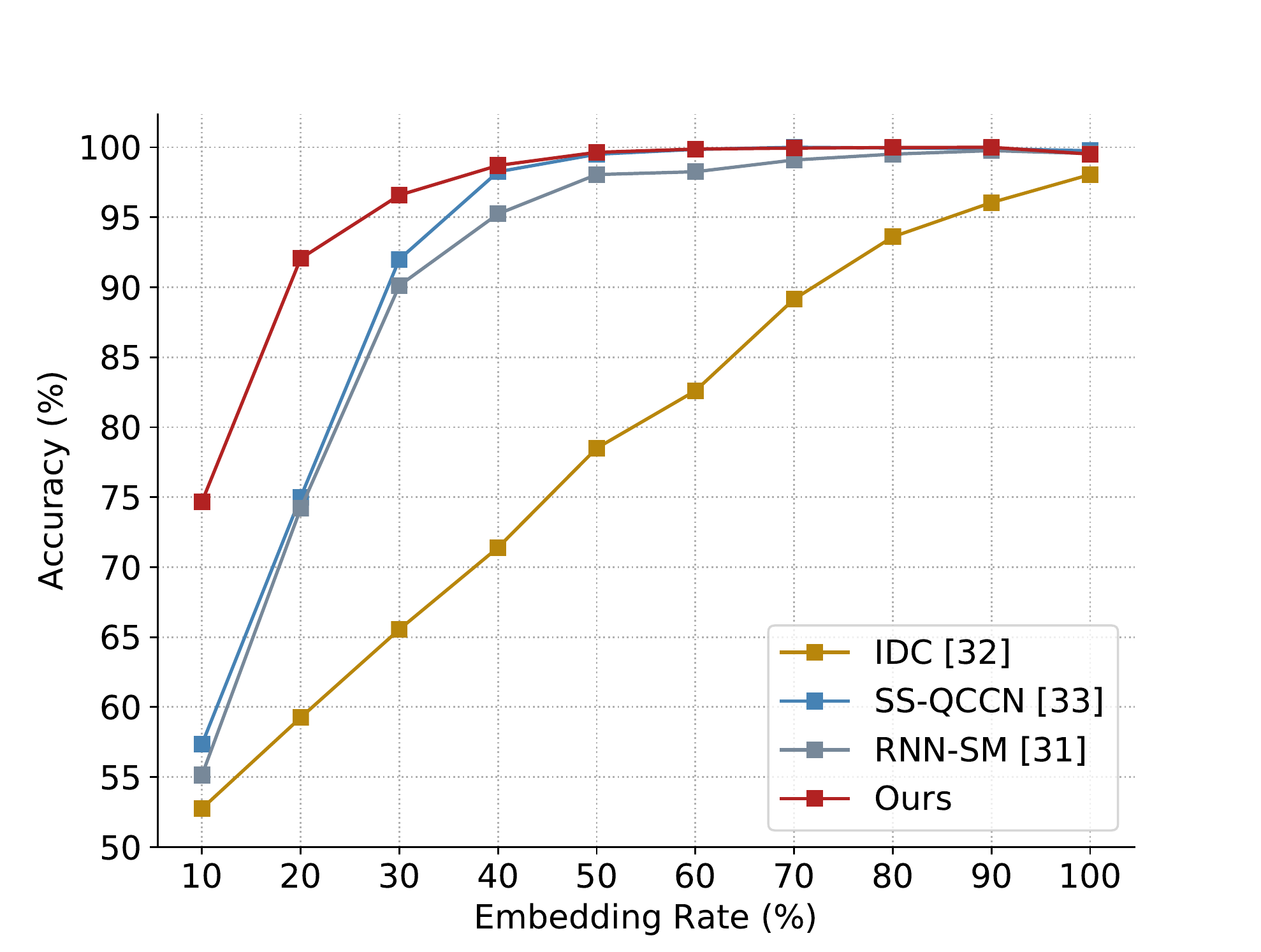}
\end{minipage}%
}%
\centering
\caption{For a speech with a length of 10s, the detection accuracy of each model varies with the embedding rate.}
\end{figure*}

\begin{figure*}[!ht]
\centering
\subfigure[ER = 10\%]{
\begin{minipage}[t]{0.19\linewidth}
\centering
\includegraphics[width=\linewidth]{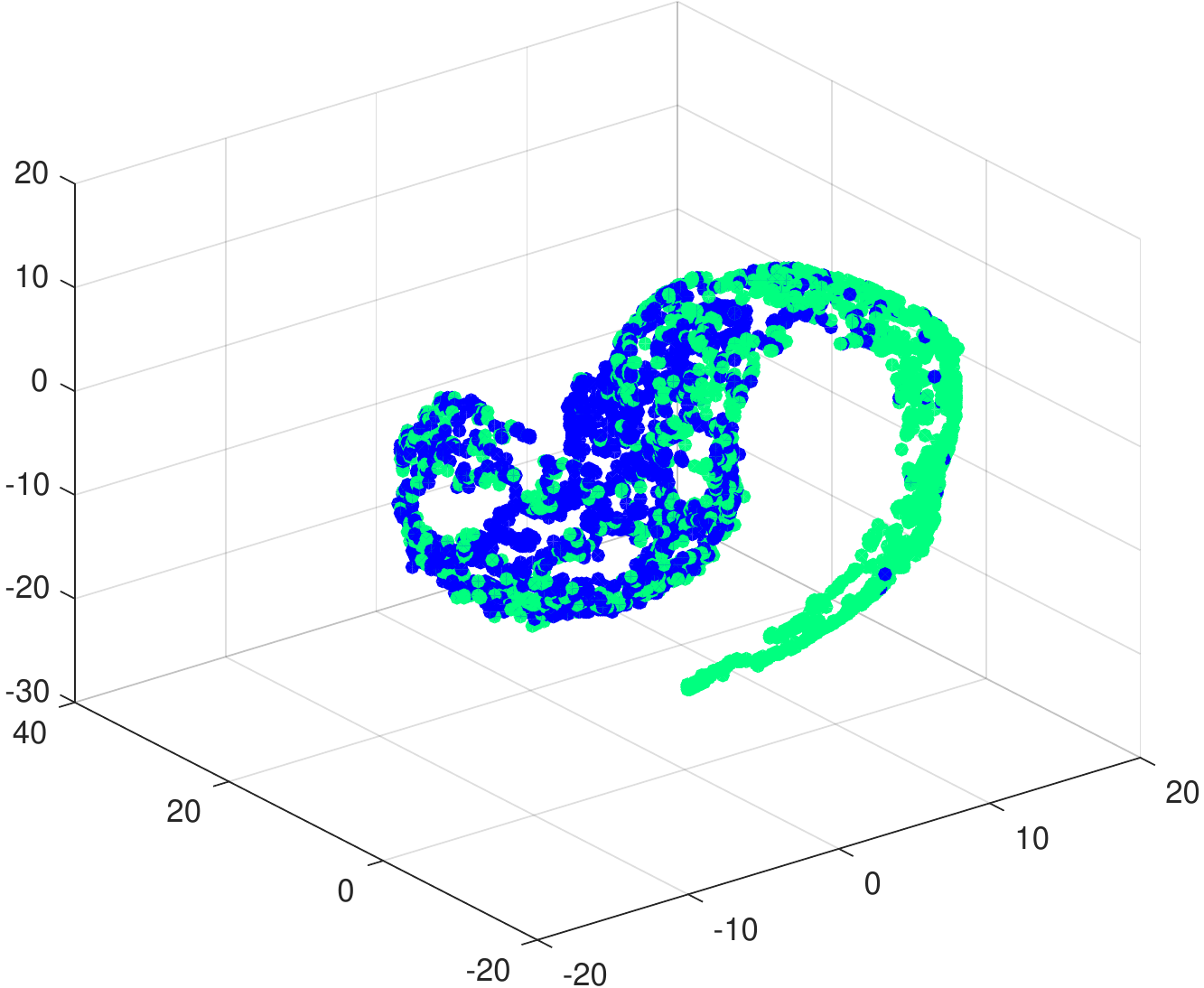}
\end{minipage}%
}%
\subfigure[ER = 20\%]{
\begin{minipage}[t]{0.19\linewidth}
\centering
\includegraphics[width=\linewidth]{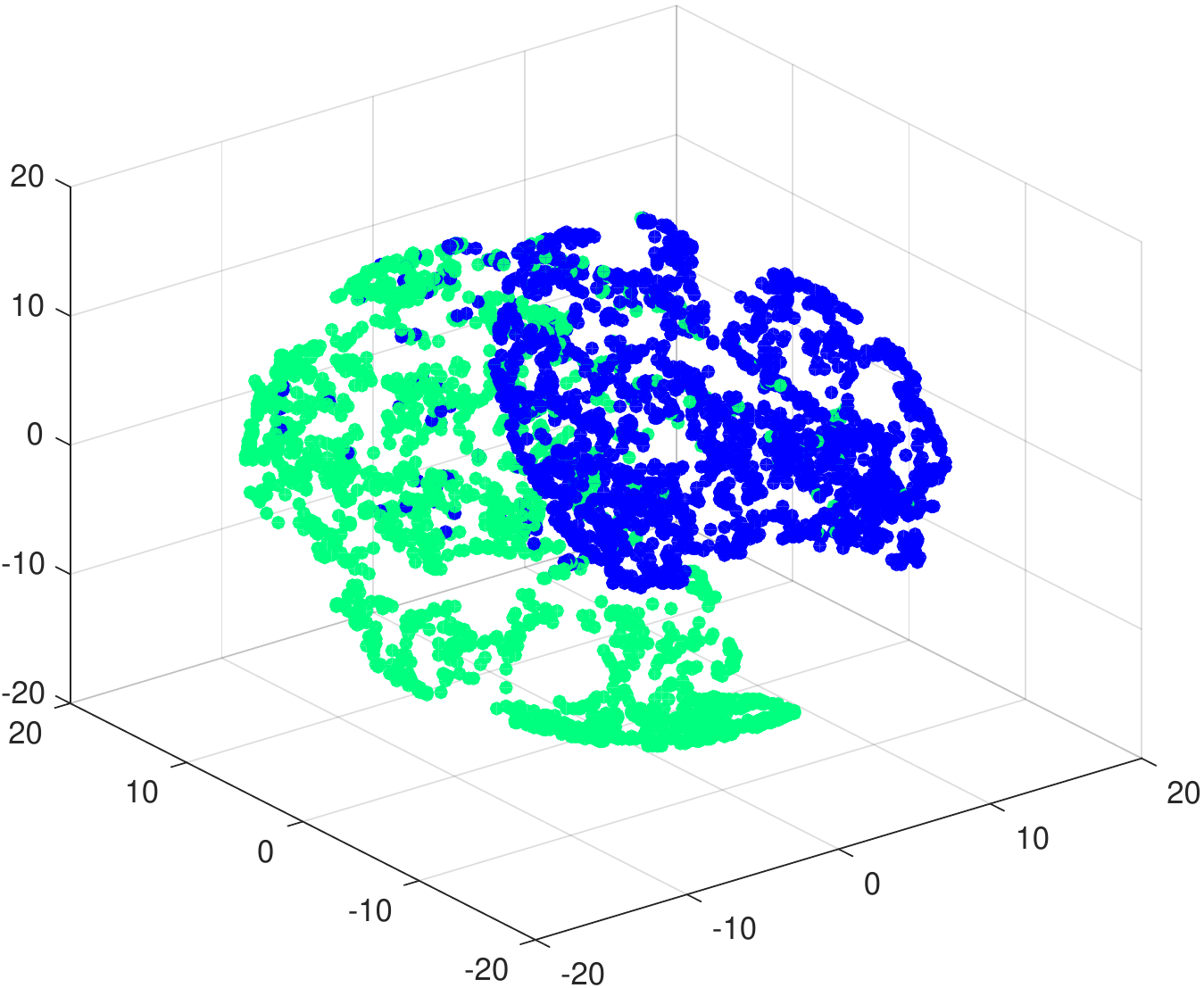}
\end{minipage}%
}%
\subfigure[ER = 30\%]{
\begin{minipage}[t]{0.19\linewidth}
\centering
\includegraphics[width=\linewidth]{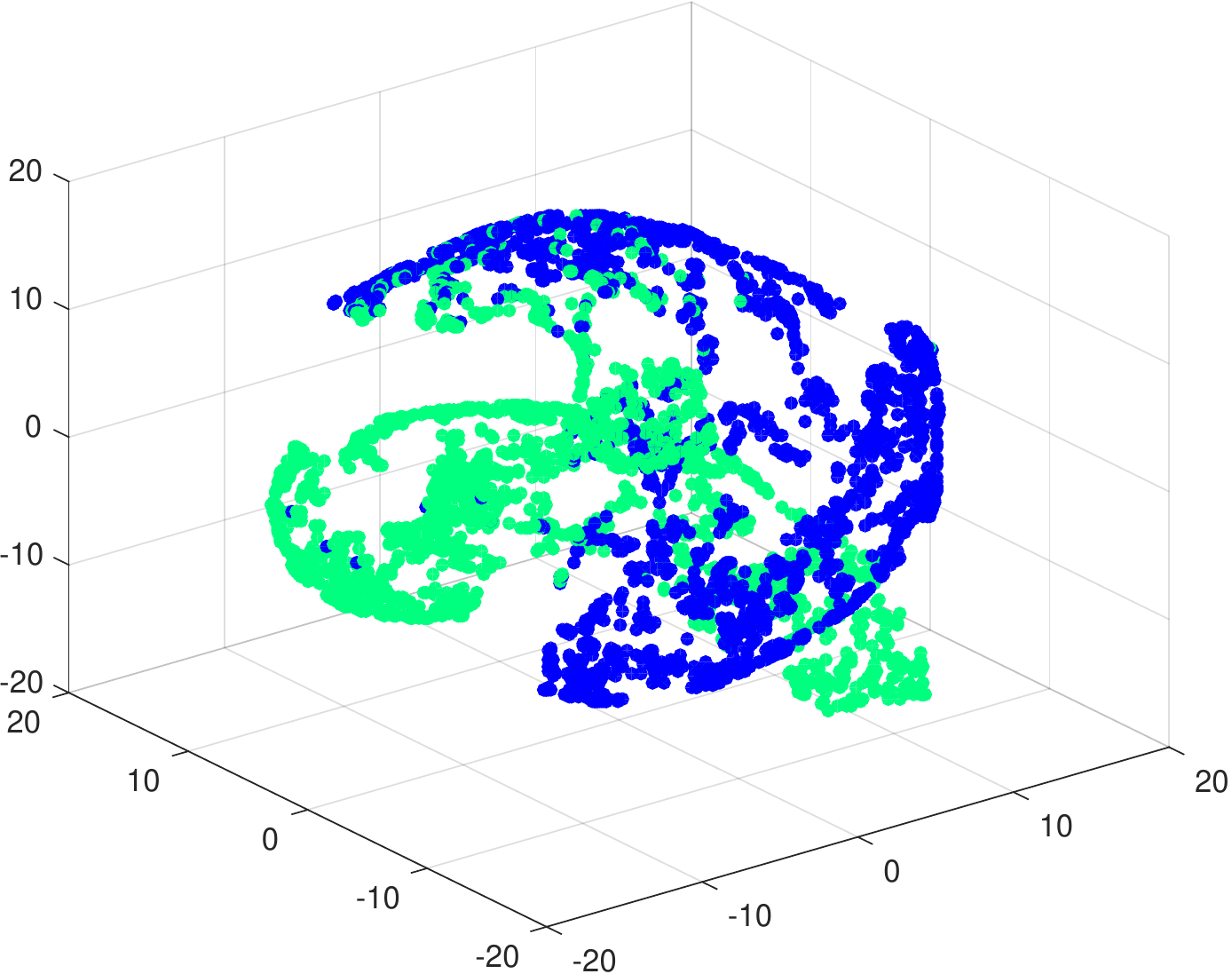}
\end{minipage}
}%
\subfigure[ER = 40\%]{
\begin{minipage}[t]{0.19\linewidth}
\centering
\includegraphics[width=\linewidth]{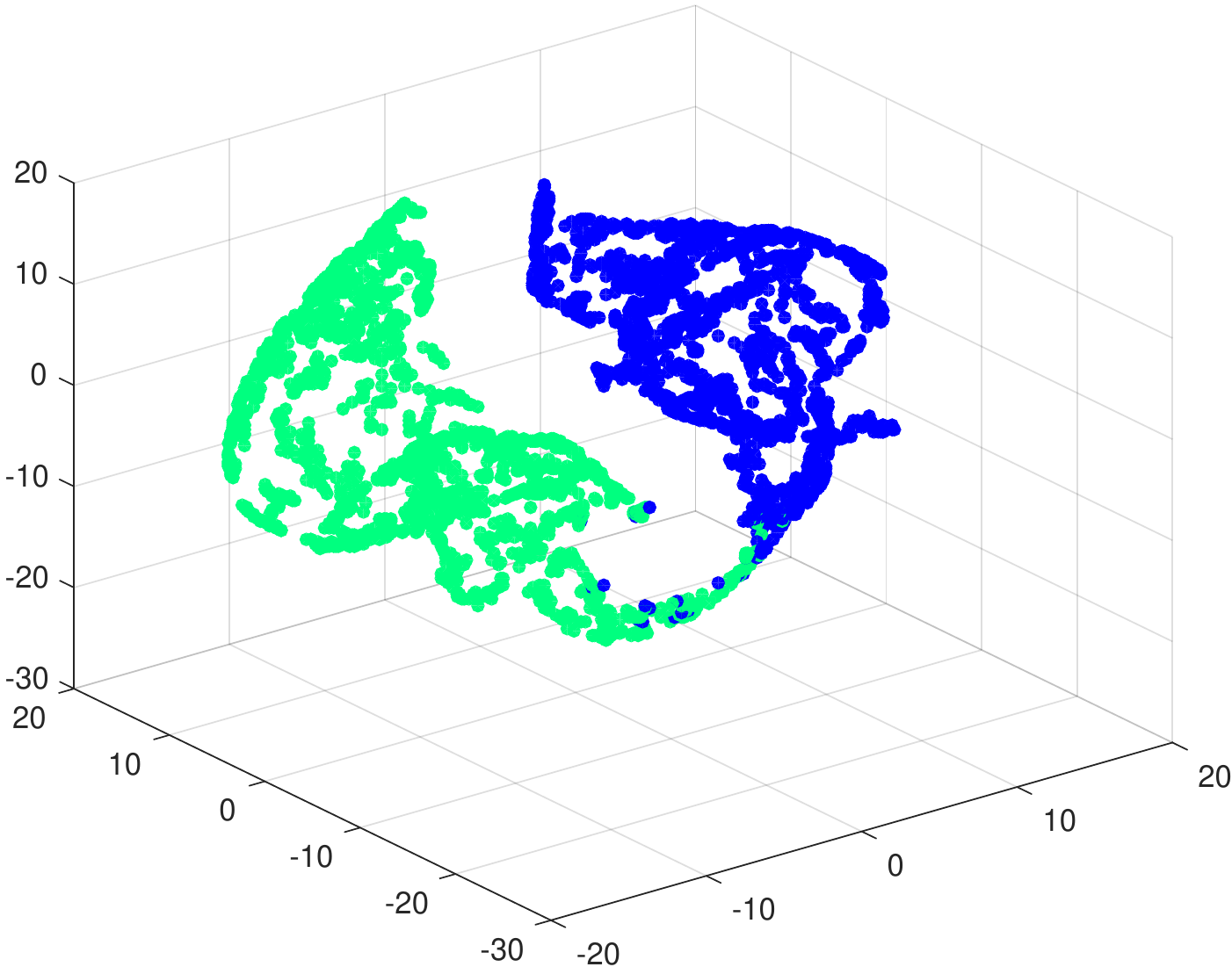}
\end{minipage}
}%
\subfigure[ER = 50\%]{
\begin{minipage}[t]{0.19\linewidth}
\centering
\includegraphics[width=\linewidth]{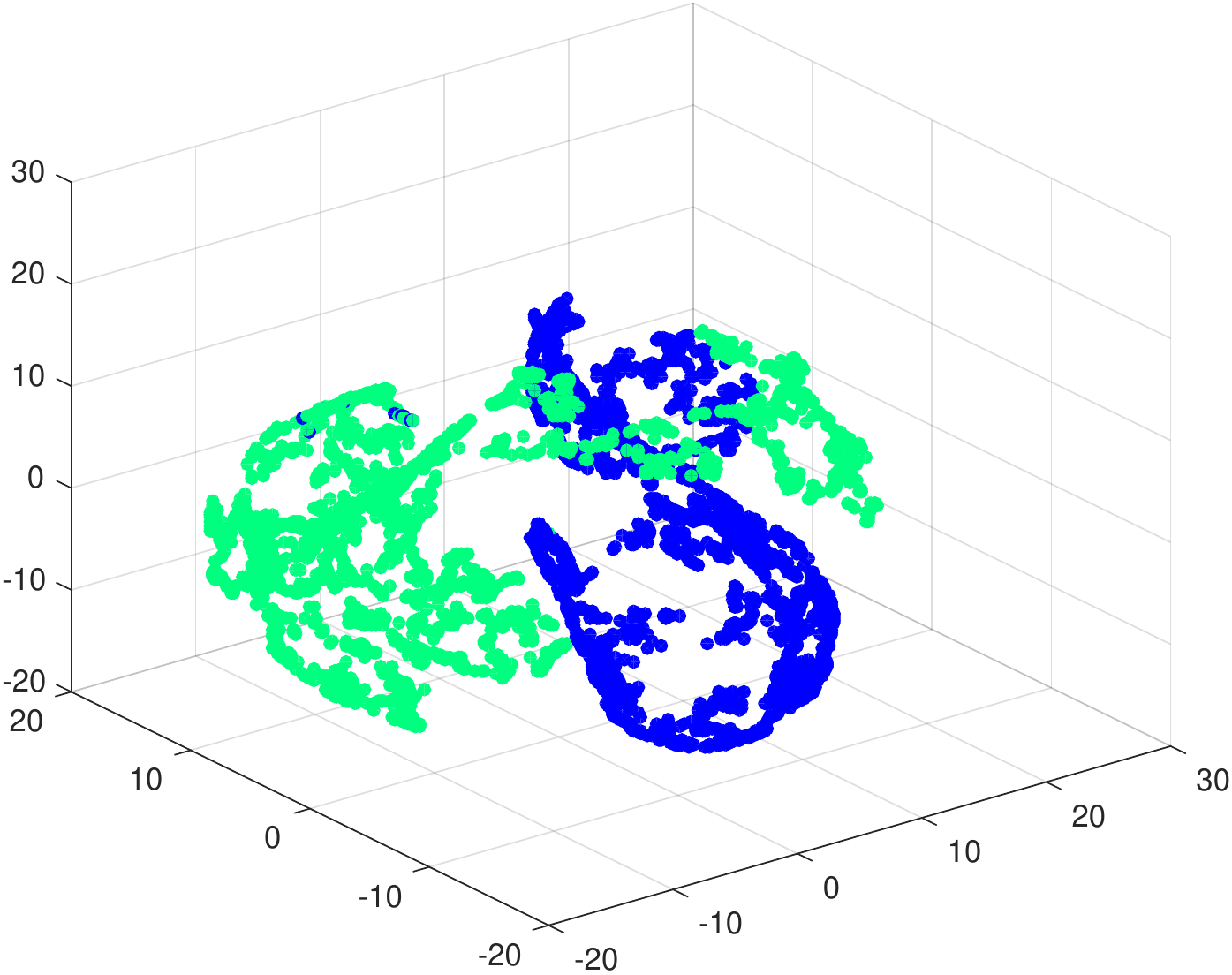}
\end{minipage}
}%
\quad

\subfigure[ER = 60\%]{
\begin{minipage}[t]{0.19\linewidth}
\centering
\includegraphics[width=\linewidth]{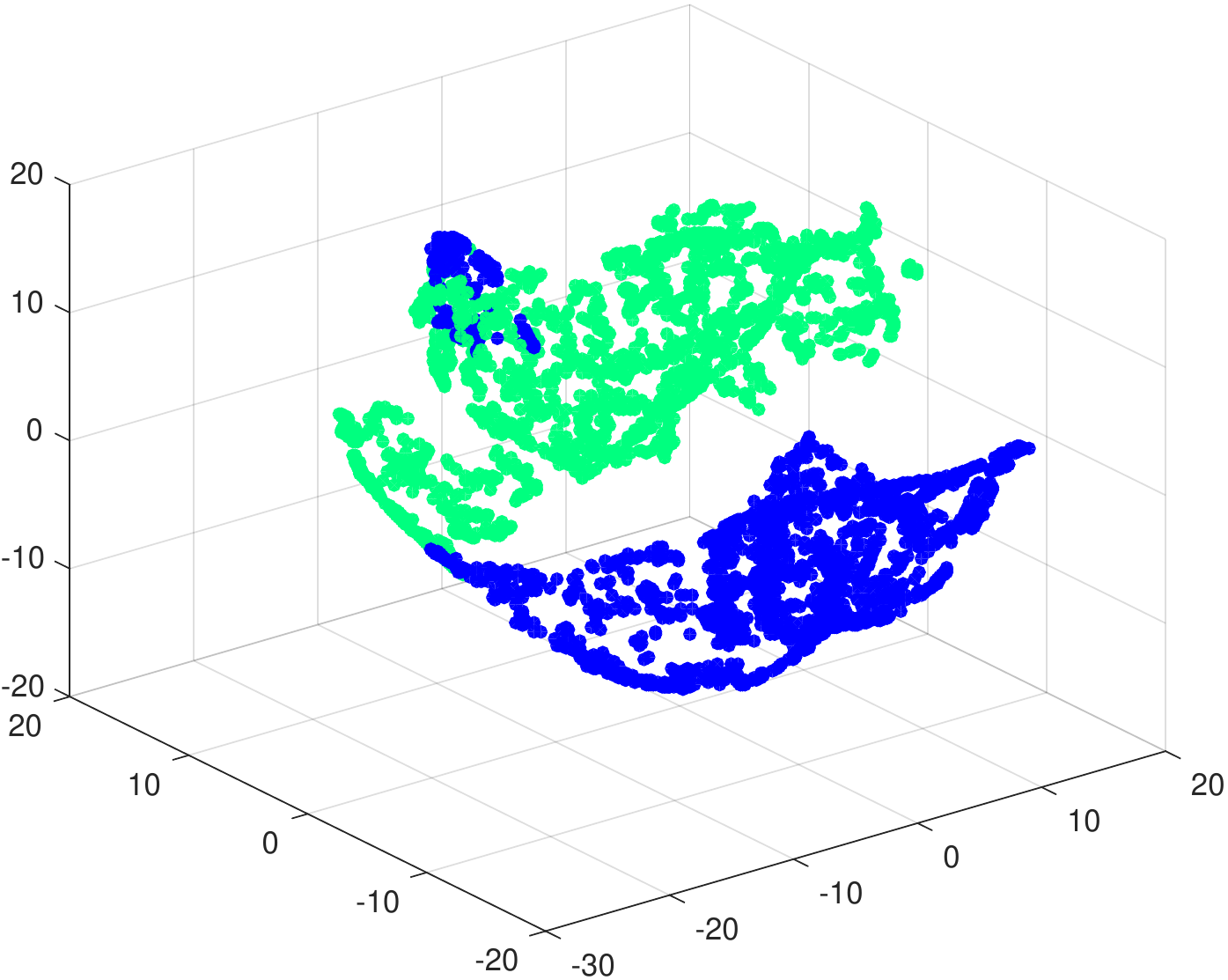}
\end{minipage}%
}%
\subfigure[ER = 70\%]{
\begin{minipage}[t]{0.19\linewidth}
\centering
\includegraphics[width=\linewidth]{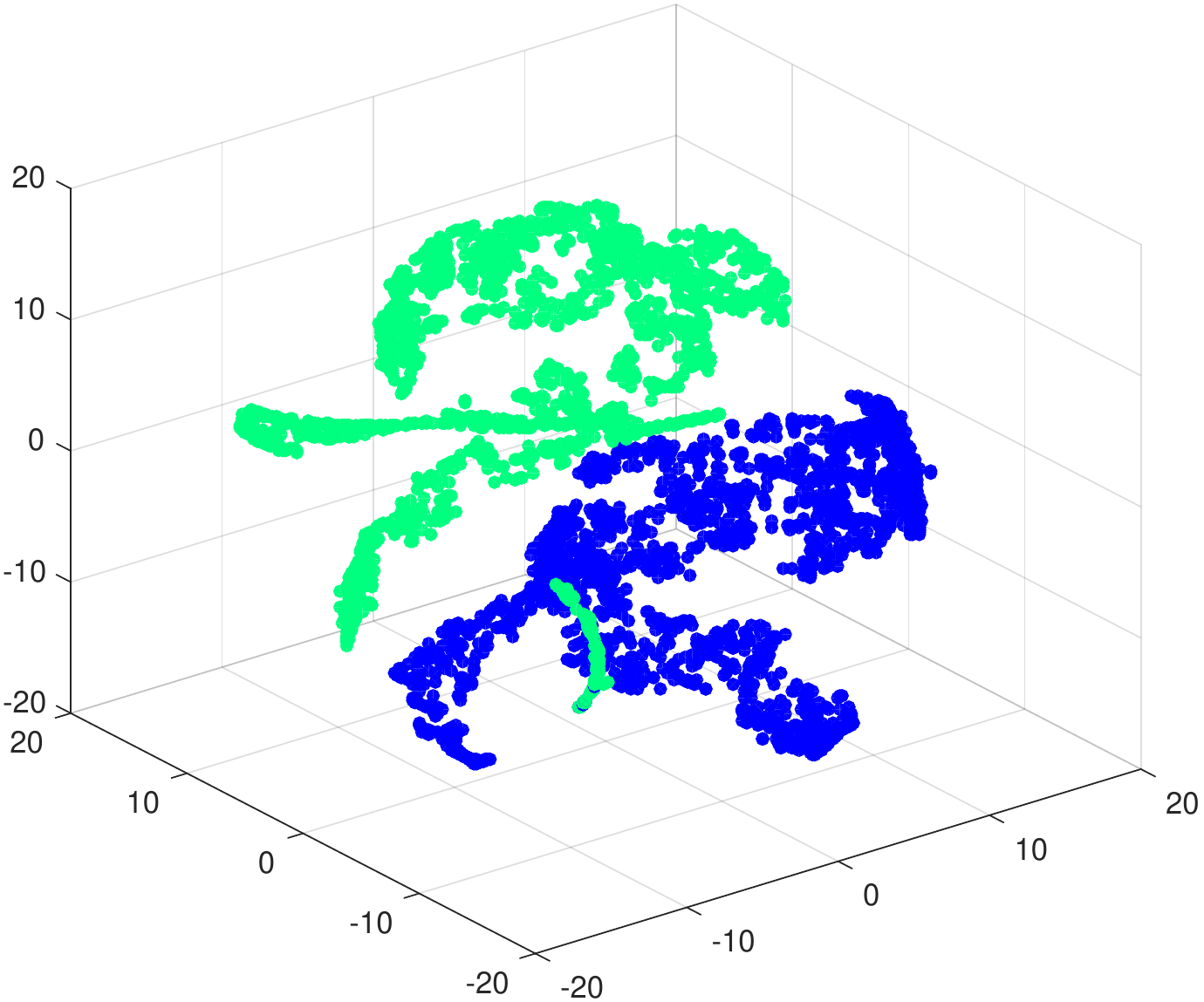}
\end{minipage}%
}%
\subfigure[ER = 80\%]{
\begin{minipage}[t]{0.19\linewidth}
\centering
\includegraphics[width=\linewidth]{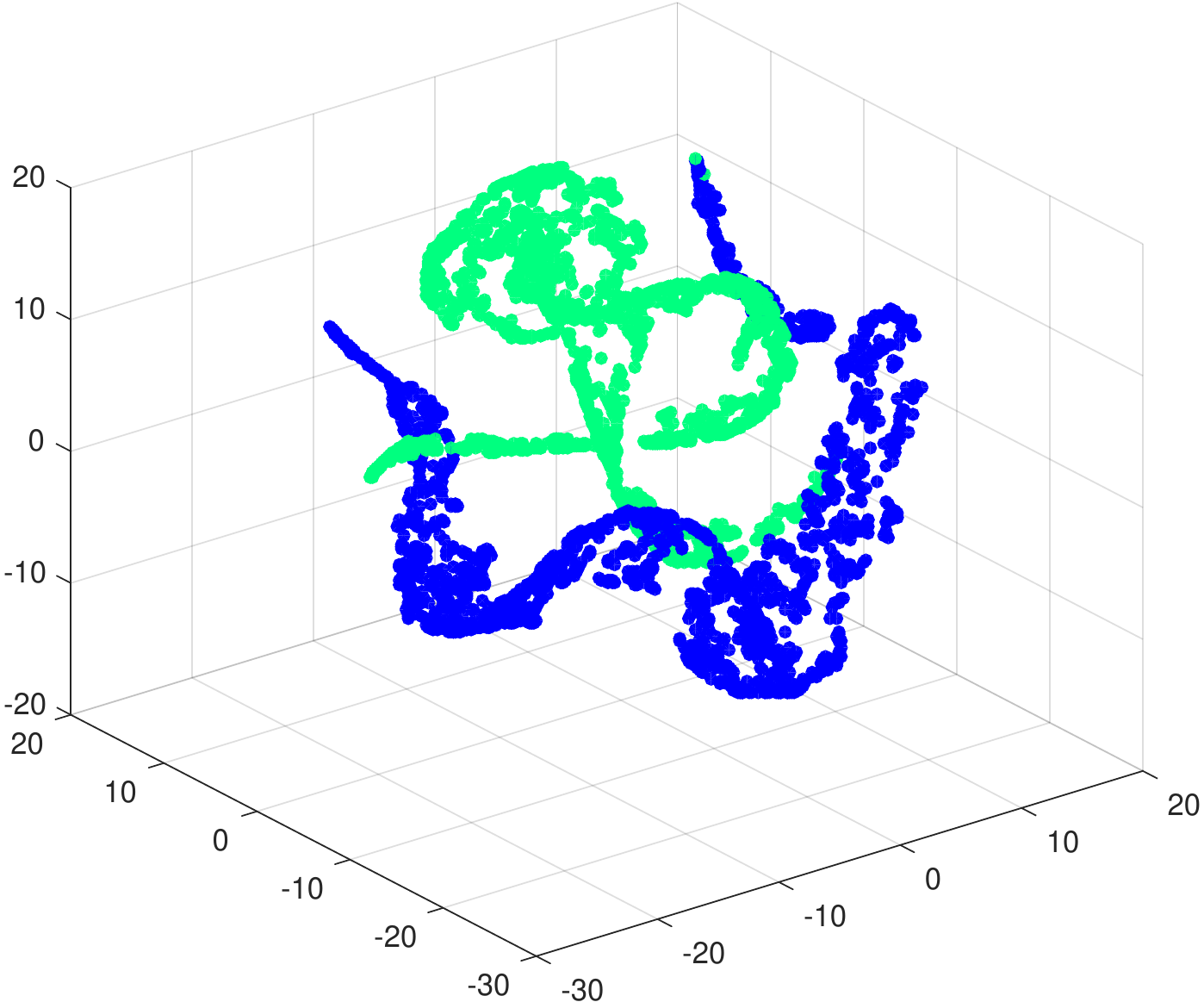}
\end{minipage}
}%
\subfigure[ER = 90\%]{
\begin{minipage}[t]{0.19\linewidth}
\centering
\includegraphics[width=\linewidth]{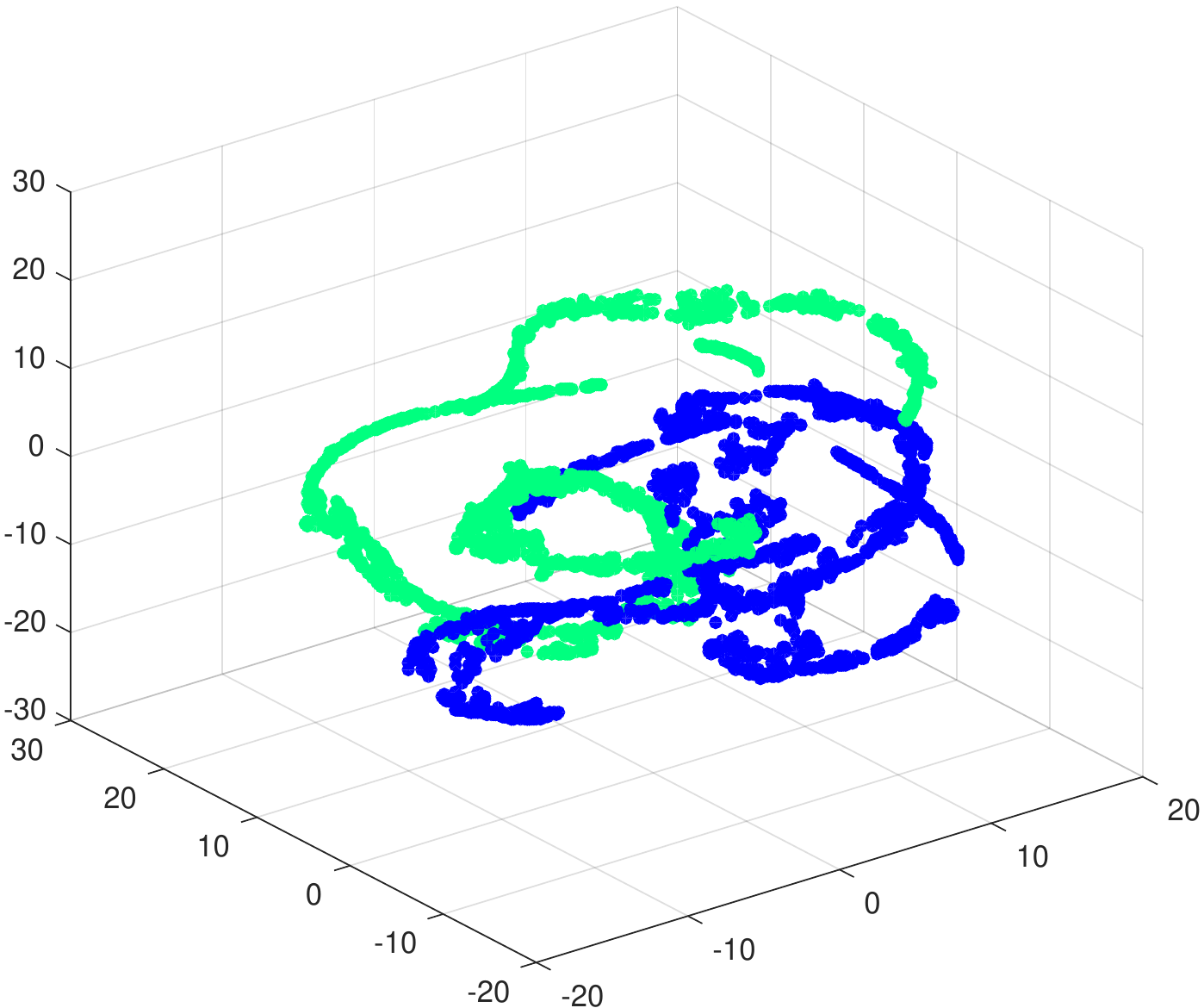}
\end{minipage}
}%
\subfigure[ER = 100\%]{
\begin{minipage}[t]{0.19\linewidth}
\centering
\includegraphics[width=\linewidth]{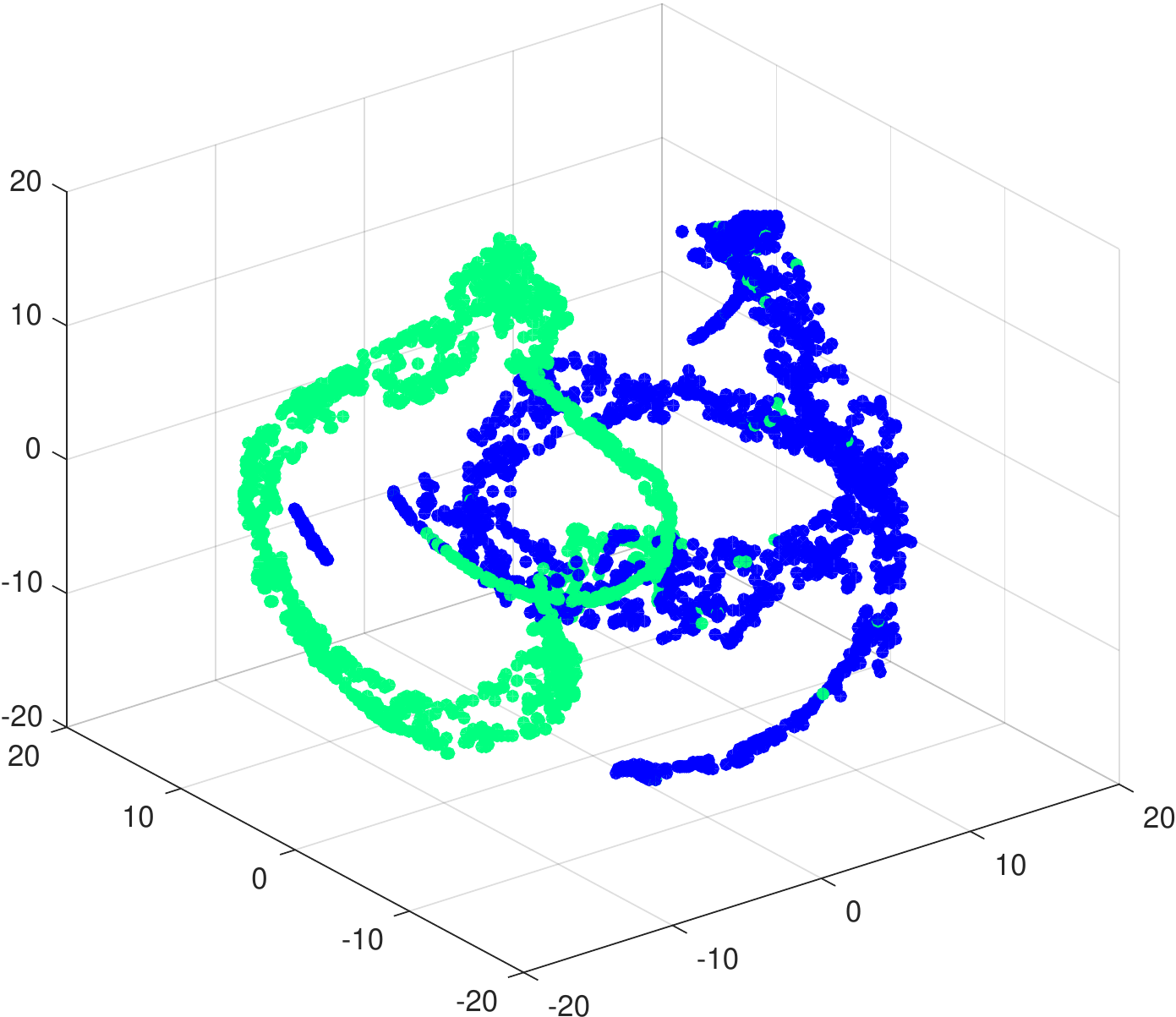}
\end{minipage}
}%

\centering
\caption{The distribution of speech features extracted by the proposed model in statistical space varies with the embedding rate. Each of the dots represents a speech with a length of 10s, the blue points indicate cover speeches, and the green points indicate stego speeches with different embedding rate concealment information. We can clearly see that as the embedding rate increases, these speech features gradually separate in the statistical space. When the embedding rate is greater than 40\%, the boundary can be clearly seen from the feature space.}
\end{figure*}

Finally, to synthesize the performance of all aspects of the model, the structure and hyper-parameters of the proposed model are setted as follows. We use three detection channels with corresponding sliding window lengths of $1$, $3$, and $5$, respectively. The channel with length $1$ mainly extracts intra-frame correlation, and the latter two channels mainly extract inter-frame correlation in different neighborhoods. The first convolutional layer in each sliding window contains $128$ convolution kernels. The width of the second layer of convolution kernels in three different channels is $3$, $5$, and $7$, and each contains $64$ convolution kernels . The size of skip matrix $S^k$ is setted to be $S^k \in \mathbb{R}^{64 \times 3}$. The dimension of the spliced feature vector $Z$ is $448$. After the feature fusion of the fully connected layer, the dimension of the feature vector $O$ is $64$. During model training, in order to strengthen the regularization and prevent overfitting, we adopt the dropout mechanism \cite{Srivastava2014Dropout} during the training process. We choose Adam \cite{Kingma2014Adam} as the optimization method. The learning rates are initially set as 0.001 and batch size is set as 256, dropout rate is 0.5.

\subsection{Evaluation Results and Discussion}

In order to objectively reflect the performance of the proposed model, in this section, we choose three state-of-the-art VoIP steganalysis algorithms which are: IDC \cite{li2012detection} and SS-QCCN \cite{li2017steganalysis} and RNN-SM \cite{Lin2018RNN} as our baselines. IDC proposed in \cite{li2012detection} extracted the transition probability between codewords in inter frames. SS-QCCN \cite{li2017steganalysis} further took the transition probability in intra frames into consideration. Both of these two models used SVM as classifier. RNN-SM \cite{Lin2018RNN} used a 2-layer Recurrent Neural Networks (RNNs) with Long Short Time Memory (LSTM) units to extract the codeword correlation features in VoIP streams and then used a feature classification model to classify those correlation features into cover speech and stego speech categories.

\subsubsection{Performance Under Different Embedding Rates}

Embedding rate (ER) is defined as the ratio of the number of embedded bits to the whole embedding capacity. In general, when the embedding rate is small, the statistical distribution of the carrier before and after steganography is small, making it easier to satisfy formula (3) and more difficult to be detected. In reality, Alice and Bob may spread the secret information over a long time frame to embed, thus reducing the average embedding rate of information to ensure the concealment of their communication. Therefore, effective steganographic detection of VoIP speech signals at low embedding rates has long been a very challenging but also very realistic research goal. We tested the detection performance of each model at different embedding rates when the length of the speech samples is 10s and the results are shown in Table 2. According to the results in Table 2, we can draw the following conclusions. Firstly, as the embedding rate increases, the detection accuracy of each models increase, which is consistent with our previous analysis. In order to visually reflect this situation, we plot the detection accuracy of each model at different embedding rates, as shown in Figure 8. Compared to other VoIP steganalysis methods, the proposed models has achieved the best detection performance in most cases, including different language and different embedding rates. 

Secondly, we find that when the embedding rate is relatively high (for example, 80\%), existing models can basically detect the steganography of VoIP speech signals effectively, and the accuracy rate can reach more than 90\%. However, when the embedding rate is very low, such as only 10\%, existing models show unsatisfactory detection performance, and the proposed model shows great advantages over other models (English: Ours (83.48\%) VS RNN-SM \cite{Lin2018RNN} (59.64\%), Chinese: Ours (77.18\%) VS RNN-SM \cite{Lin2018RNN} (59.14\%)). These results prove that the proposed model can extract the statistical distribution difference of VoIP speech signal features before and after steganography more effectively than other methods. Even when the embedding rate is very low and the difference of statistical features is not obvious, we can still detect with high accuracy. We use t-Distributed Stochastic Neighbor Embedding (t-SNE)\cite{Maaten2014Accelerating} technology to reduce and visualize the VoIP speech signal features (vector $O$ in Equation (21)) extracted by the proposed model. The results are shown in Figure 9. In Figure 9, each point represents an input VoIP speech signal, the blue points indicate cover speeches, and the green points indicate stego points. From Fig. 9, we can find that when the embedding rate is only 10\%, the cover speech and the stego speech have a large overlapping area in the feature space, which makes steganalysis models difficult to distinguish them. However, as the embedding rate increases, the distribution of the cover speech and the stego speech in the feature space is gradually separated. When the embedding rate is greater than 40\%, the boundary can be clearly seen from the feature space. The results in Figure 9 intuitively reflect the proposed model's ability to extract and analyze the features of steganographic speech at different embedding rates.

Thirdly, from Table 2, we also notice that, in most of the cases, the accuracy of English speeches are higher than that of Chinese speech samples when they have the same embedding rate. This phenomenon may be explained by the different characteristics of the two languages. English is composed by 20 vowels and 28 consonants. However, in Chinese, there are 412 kinds of syllables. The diversity makes the correlations between codewords in Chinese more complicated, and it is therefore more difficult to perform steganographic detection.

\subsubsection{Performance Under Different Clip Length}

\begin{table*}[!ht]
\centering
\caption{\label{tab:4}When the embedding rate is 100\%, the detection results of each model under different speech length.}
\resizebox{\textwidth}{50mm}{
\begin{tabular}{c|c|c|cccccccccc}
\toprule[2pt]
\multirow{2}{*}{Language} &\multirow{2}{*}{Method} &\multirow{2}{*}{Metric} &\multicolumn{10}{|c}{Sample Length (s)}\\

 & & &0.1 &0.2 &0.3 &0.4 &0.5 &0.6 &0.7 &0.8 &0.9 &1.0\\
\hline
\multirow{12}{*}{English} &\multirow{3}{*}{IDC \cite{li2012detection}} &Acc (\%) &85.40 & 88.00 & 88.50 & 89.25 & 90.10 & 91.45 & 91.40 & 92.40 & 92.95 & 93.70\\
& &FP (\%) &14.60 &11.40 &10.70 &11.60 &8.90 &8.00 &8.30 &8.60 &7.60 &6.10\\
& &FN (\%)&14.60 &12.60 &12.30 &9.90 &10.90 &9.10 &8.90 &6.60 &6.50 &6.50\\
\cline{2-13}
&\multirow{3}{*}{SS-QCCN \cite{li2017steganalysis}} &ACC (\%) &82.00 &88.85 &92.15 &95.00 &95.70 &96.15 &96.25 &96.90 &96.90 &98.00\\
& &FP (\%) &8.00 &8.90 &5.80 &4.60 &4.20 &2.90 &3.70 &1.80 &1.70 &0.60\\
& &FN (\%) &28.00 &13.40 &9.90 &5.40 &4.40 &4.80 &3.80 &4.40 &4.50 &3.40\\
\cline{2-13}
&\multirow{3}{*}{RNN-SM \cite{Lin2018RNN}} &ACC (\%) &90.40 &95.50 &97.38 &97.81 &98.16 &98.23 &98.38 &\textbf{98.48} &\textbf{98.49} &\textbf{98.54}\\
& &FP (\%) &10.02 &4.66 &2.46 &1.77 &1.57 &1.76 &1.06 &1.04 &0.85 &0.81\\
& &FN (\%) &8.16 &3.53 &3.48 &2.80 &2.24 &2.01 &1.91 &1.59 &1.04 &1.47\\
\cline{2-13}
&\multirow{3}{*}{Ours} &ACC (\%) &\textbf{91.59} &\textbf{95.63} &\textbf{97.40} &\textbf{97.85} &\textbf{98.21} &\textbf{98.36} &\textbf{98.40} &98.43 &\textbf{98.49} &98.47\\
& &FP (\%) &2.06 &1.39  &3.71  &2.08  &1.83  &2.29  &2.37  &2.98  &2.42  &0.33\\
& &FN (\%) &14.76  &7.35  &1.49  &2.22  &1.75  &0.99  &0.83  &0.16  &0.60  &2.73\\
\hline
\hline
\multirow{12}{*}{Chinese} &\multirow{3}{*}{IDC \cite{li2012detection}} &ACC (\%) &86.80 &88.65 &90.20 &90.50 &91.20 &92.25 &93.10 &94.25 &94.70 &94.05\\
& &FP (\%) &13.60 &11.40 &9.40 &9.30 &9.90 &7.80 &6.70 &6.70 &6.10 &7.10\\
& &FN (\%) &12.80 &11.30 &10.20 &9.70 &7.70 &7.70 &7.10 &4.80 &4.50 &4.80\\
\cline{2-13}
&\multirow{3}{*}{SS-QCCN \cite{li2017steganalysis}} &ACC (\%) &81.20 &90.05 &93.75 &95.25 &96.50 &97.45 &97.60 &98.30 &98.10 &98.50\\
& &FP (\%) &9.50 &8.30 &7.10 &4.80 &3.40 &1.90 &2.70 &1.80 &1.40 &0.90\\
& &FN (\%) &28.10 &11.60 &5.40 &4.70 &3.60 &3.20 &2.10 &1.60 &2.40 &2.10\\
\cline{2-13}
&\multirow{3}{*}{RNN-SM \cite{Lin2018RNN}} &ACC (\%) &90.91 &95.91 &97.03 &97.72 &98.09 &98.12 &98.51 &98.69 &99.06 &98.86\\
& &FP (\%) &10.02 &4.66 &2.46 &1.77 &1.57 &1.76 &1.06 &1.04 &0.85 &0.81\\
& &FN (\%) &8.16 &3.53 &3.48 &2.80 &2.24 &2.01 &1.91 &1.59 &1.04 &1.47\\
\cline{2-13}
&\multirow{3}{*}{Ours} &ACC (\%) &\textbf{91.84} &\textbf{96.12} &\textbf{97.70} &\textbf{98.32} &\textbf{98.56} &\textbf{98.40} &\textbf{98.99} &\textbf{98.80} &\textbf{99.13} &\textbf{98.95}\\
& &FP (\%) &7.25 &2.79  &2.50  &0.95  &3.43  &1.88  &0.98  &1.97  &0.94  &0.33\\
& &FN (\%) &9.07 &4.97  &2.10  &2.41  &0.00  &1.32  &1.04  &0.43 & 0.80  &1.77\\
\bottomrule[2pt] 
\end{tabular}}
\end{table*}

\begin{table}[!h]
\centering
\caption{\label{tab:4}The detection accuracy of each model for a speech signal with a short length and a low embedding rate.}

\begin{tabular}{c|c|c|cccccccccc}
\toprule[2pt]
\multirow{2}{*}{length} &\multirow{2}{*}{language} &\multirow{2}{*}{Method} &\multicolumn{4}{|c}{Embedding Rate}\\

 & & &10\% &20\% &30\% &40\%\\
\hline
\multirow{8}{*}{0.1s} &\multirow{4}{*}{EN} &IDC \cite{li2012detection} &52.95 &57.65  &62.90 &67.05\\
& &SS-QCCN \cite{li2017steganalysis} &50.55 &54.80 &58.25  &59.05\\
& &RNN-SM \cite{Lin2018RNN}&55.39 & 60.25 & 67.43 & 70.28\\
& &Ours &\textbf{55.99} & \textbf{62.44} & \textbf{67.58} & \textbf{72.36}\\
\cline{2-7}
 &\multirow{4}{*}{CN} &IDC \cite{li2012detection} &53.90 & 58.85 & 63.70 & 68.05\\
& &SS-QCCN \cite{li2017steganalysis} &51.55 & 54.80 & 58.25 & 59.05\\
& &RNN-SM \cite{Lin2018RNN}&54.71 & 60.48 & 63.60 & 68.18\\
& &Ours &\textbf{55.20} & \textbf{61.71} & \textbf{67.15} & \textbf{72.04}\\
\hline
\hline
\multirow{8}{*}{0.3s} &\multirow{4}{*}{EN} &IDC \cite{li2012detection} &54.55 & 58.15 & 63.65 & 69.50\\
& &SS-QCCN \cite{li2017steganalysis}&53.15 & 58.25 & 62.90 & 71.45\\
& &RNN-SM \cite{Lin2018RNN}&59.68 & 70.05 & 77.17 & 77.27\\
& &Ours &\textbf{61.23} & \textbf{70.61} & \textbf{78.21} & \textbf{84.43}\\
\cline{2-7}
 &\multirow{4}{*}{CN} &IDC \cite{li2012detection} &54.50 & 60.10 & 65.70 & 70.05\\
& &SS-QCCN \cite{li2017steganalysis} &53.15 & 58.25 & 62.90 & 71.45\\
& &RNN-SM \cite{Lin2018RNN}&57.61 & 66.81 & 74.60 & 80.08\\
& &Ours &\textbf{59.66} & \textbf{70.02} & \textbf{77.87} & \textbf{82.75}\\
\hline
\hline
\multirow{8}{*}{0.5s} &\multirow{4}{*}{EN} &IDC \cite{li2012detection} &52.00 & 60.05 & 64.00 & 69.20\\
& &SS-QCCN \cite{li2017steganalysis} &54.00 & 60.95 & 67.00 & 75.15\\
& &RNN-SM \cite{Lin2018RNN}&62.00 & 73.49 & 80.00 & 87.98\\
& &Ours &\textbf{64.33} & \textbf{75.72} & \textbf{83.80} & \textbf{89.49}\\
\cline{2-7}
 &\multirow{4}{*}{CN} &IDC \cite{li2012detection} &57.00 & 60.00 & 65.00 & 71.30\\
& &SS-QCCN \cite{li2017steganalysis} &54.00 & 62.00 & 67.00 & 75.65\\
& &RNN-SM \cite{Lin2018RNN}&\textbf{71.00} & 71.00 & 78.00 & 85.75\\
& &Ours &69.45 & \textbf{80.21} & \textbf{82.19} & \textbf{86.83}\\
\hline
\hline
\multirow{8}{*}{1.0s} &\multirow{4}{*}{EN} &IDC \cite{li2012detection} &57.00 & 65.00 & 72.00 & 78.00\\
& &SS-QCCN \cite{li2017steganalysis} &55.00 & 64.00 & 73.00 & 82.00\\
& &RNN-SM \cite{Lin2018RNN}&68.00 & 69.00 & 86.00 & 91.00\\
& &Ours &\textbf{69.09} & \textbf{81.36} & \textbf{88.91} & \textbf{93.18}\\
\cline{2-7}
 &\multirow{4}{*}{CN} &IDC \cite{li2012detection} &55.00 & 64.00 & 69.00 & 77.00\\
& &SS-QCCN \cite{li2017steganalysis} &55.00 & 64.00 & 73.00 & 82.00\\
& &RNN-SM \cite{Lin2018RNN}&62.00 & 75.00 & 84.00 & 89.00\\
& &Ours &\textbf{66.21} & \textbf{78.94} & \textbf{87.20} & \textbf{92.41}\\
\bottomrule[2pt] 
\end{tabular}
\end{table}

Compared with the steganalysis methods based on static carrier, since VoIP speech signals are transmitted online in real time, we usually require VoIP steganalysis to achieve sufficiently high detection accuracy in a short enough time. Therefore, we tested the performance of each model for different speech lengths. Table 3 shows the detection performance of each model for different length speeches when the embedding rate is 100\%. From the results in Table 3, firstly, we notice that as the length of speech changes from short to long, the detection accuracy of each model is gradually improved. The reason might be that there are codeword correlations between various distance frames in the speech signal. In general, the correlations between nearer frames are stronger than the correlations between distant frames. Therefore, once the covert information is embedded in the speech signal, it first effects the correlations between the distant frames, and which are easier to be found in the longer speech signal. The shorter speech signals have stronger codeword correlations and are less likely to be corrupted by the embedded secret information, so they are more difficult to detect. Secondly, the experimental results also show that the detection performance of the proposed model is superior to other models for most cases. In particular, when the length is short, for example, only 0.1 seconds, the proposed method can achieve 91.59\% (English) and 91.84\% (Chinese) of the detection accuracy and exceed all previous models. This means that if Alice and Bob are making covert communication using VoIP, Eve can have more than 91\% confidence in whether they are transmitting secret information within 0.1 second after their call start. This will further enhance Eve's maintenance of network security, which has very important practical significance.

\begin{table*}[!ht]
\centering
\caption{\label{tab:4}The detection time (ms) spent by each model for different lengths of speech.}
\resizebox{\textwidth}{15mm}{
\begin{tabular}{c|c|cccccccccccccccc}
\toprule[1.5pt]
\multirow{2}{*}{Device} &\multirow{2}{*}{Method} &\multicolumn{16}{|c}{Sample Length (s)}\\

 & &0.1 &0.2 &0.3 &0.4 &0.5 &0.6 &0.7 &0.8 &0.9 &1.0 &2.0 &3.0 &4.0 &5.0 &6.0 &10.0\\
\hline
\multirow{2}{*}{RNN-SM \cite{Lin2018RNN}} &Mean &0.591 &1.006 &1.418 &1.801 &2.239 &2.614 &3.0305 &3.429 &3.731 &4.183 &8.093 &12.269 &15.740 &19.700 &23.038 &38.645\\
&sd &0.125  &0.170 &0.190 &0.209 &0.266 &0.264 &0.441 &0.431 &0.370 &0.497 &0.954 &2.058 &2.216 &2.396 &2.305 &3.595\\
\hline
\multirow{2}{*}{Ours} &Mean &\textbf{0.519}  &\textbf{0.550} &\textbf{0.579}  &\textbf{0.585} &\textbf{0.613} &\textbf{0.644} &\textbf{0.673} &\textbf{0.672} &\textbf{0.702} &\textbf{0.733} &\textbf{0.992} &\textbf{1.200} &\textbf{1.400} &\textbf{1.652}  &\textbf{1.911} &\textbf{2.876}\\
&sd &0.147  &0.160 &0.167 &0.170 &0.174 &0.183 &0.192 &0.180 &0.207 &0.213 &0.275 &0.347 &0.430 &0.464 &0.528 &0.784\\
\bottomrule[1.5pt] 
\end{tabular}}
\end{table*}

Combining the previous two experiments, we further tested the performance of each model in terms of short durations and low embedding rates. Table 4 lists the detection results of each method when clip length are 0.1s, 0.3s, 0.5s and 1s and embedding rate ranges from 10\% to 40\%. From Table 4 we can get the same conclusions with the previous two sets of experiments, that is, as the embedding rate and speech length increase, the detection performance of each model gradually increases. At the same time, we also notice that effective steganalysis for the VoIP speech signals with an extremely short length (less than 1 second) and low embedding rate (less than 30\%) are still very challenging. Nevertheless, our method has made remarkable progress compared with the previous methods and we think it has very important value and significance.

\subsubsection{Time Efficiency in Different Clips}

As we have analyzed before, VoIP speech signals are transmitted online in real time. Therefore, we usually require the VoIP Steganography analysis model to be efficient enough to meet real-time requirements. In the previous experiments, we tested the performance of each model for different speech signal lengths (Table 3). Here we further test the detection efficiency of the proposed model. Both IDC \cite{li2012detection} and SS-QCCN \cite{li2017steganalysis} depend on SVM algorithm and they take too much time to extract features from the VoIP speech stream. Therefore, they are not suitable for online real-time detection. The RNN-SM model proposed by Z. Lin \emph{et al.} \cite{Lin2018RNN} also has the ability for fast detection of VoIP tream, and both RNN-SM and our model are based on neural network models, thus we mainly compare the efficiency of our model with RNN-SM model. We tested the detection efficiency of these two models for different lengths of speech in the same environment. The results are shown in Table 5 and Figure 10.

As can be seen from Table 5 and Figure 10, firstly, as the length of the speech increases, the detection time required for both models increases almost linearly. Secondly, we also note that the detection time of our model increases with the sample length at a much slower rate than that of RNN-SM model. For example, when the sample length is only 0.1 seconds, there are little difference in detection time between these two models (Ours: 0.519$\pm$0.147 (ms) VS RNN-SM \cite{Lin2018RNN}: 0.591$\pm$0.125 (ms)). When the sample length increases to 10 seconds, the detection time required by RNN-SM is more than $13$ times that of the proposed model (Ours: 2.876$\pm$0.784 (ms) VS RNN-SM \cite{Lin2018RNN}: 38.645$\pm$3.595 (ms)). This is because the RNN-SM model uses a frame-by-frame iterative calculation when extracting inter-frame correlation features. In our model, we extract correlation features of all frames in a sliding window at one time, so the time cost is shorter. This part of the experiment further proves the high efficiency of the proposed model and enables almost real-time steganalysis and detection of VoIP voice signals, which has strong practical significance.

\begin{figure}[ht]
\centering
\includegraphics[height=7cm,width=\linewidth]{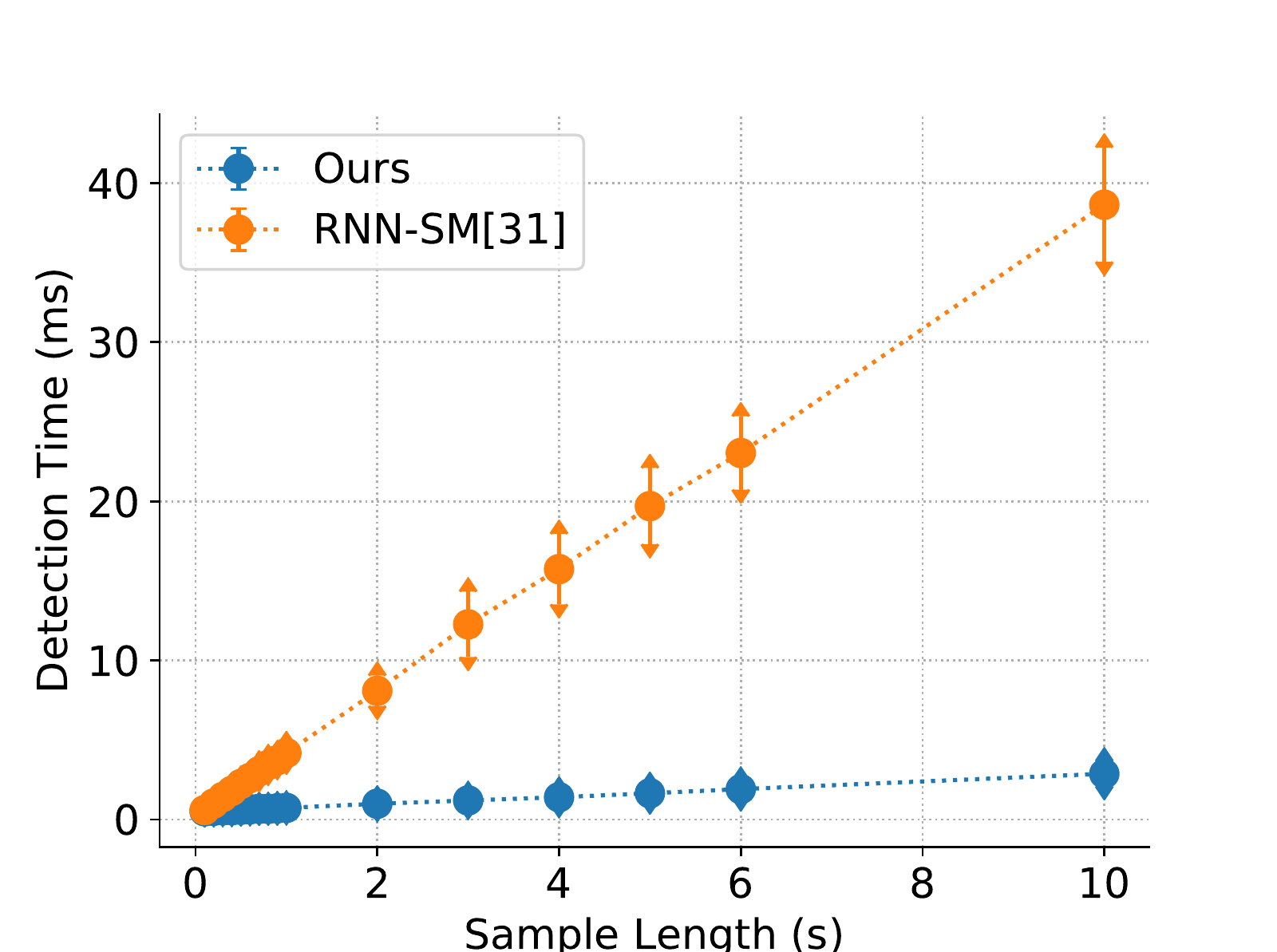}
\caption{The detection time cost by each model varies with the length of the speech.}
\label{fig:3}
\end{figure}

\section{Conclusion}

In order to solve the two major challenges in the field of VoIP steganalysis, namely: high performance and real-time detection for low embedded rate speech signals, in this paper, combined with the sliding window detection algorithm and Convolutional Neural Network (CNN), we propose a real-time VoIP steganalysis method which based on multi-channel convolutional sliding windows (CSW). It uses multi-channel sliding detection windows to extract correlations features between frames and different neighborhood frames in a VoIP signal. Within each sliding window, we design two feature extraction channels to extract both low-leavel features and high-level features of the input signal. We used a large number of experiments to verify our model in many aspects. Experimental results show that the proposed model outperforms all the previous methods, especially in the case of low embedding rate, which shows state-of-the-art performance. In addition, we also tested the detection efficiency of the proposed model, and the results show that it can achieve almost real-time detection of VoIP speech signals. We hope that this paper will serve as a reference guide for researchers to facilitate the design and implementation of better VoIP steganalysis.




%

\section*{Acknowledgment}

This research is supported by the National Key R$\&$D Program (SQ2018YGX210002) and the National Natural Science Foundation of China (No.U1536207 and No.U1636113).

\ifCLASSOPTIONcaptionsoff
  \newpage
\fi



\bibliographystyle{IEEEtran}
\end{document}